\documentclass[%
-preprint,
 amsmath,amssymb,
 aps,
]{revtex4-2}

\usepackage{graphicx}
\usepackage{dcolumn}
\usepackage{bm}
\usepackage{ulem}[normalem]

\begin{document}

\preprint{APS/123-QED}

\title{The evolution of a primordial binary black hole due to interaction\\
with cold dark matter\\
and the formation rate of gravitational wave events}

\author{Sergey Pilipenko}%
\author{Maxim Tkachev}%
\author{Pavel Ivanov}
 \email{pbi20@cam.ac.uk}
\affiliation{%
Astro Space Center, P. N. Lebedev Physical Institute of RAS, Profsojuznaya 84/32, Moscow 117997, Russia
}%

\date{\today}

\begin{abstract}
In this Paper we consider a problem of formation and evolution of orbital parameters of a binary primordial black hole (PBH) due to gravitational interaction with clustering cold dark matter (CDM). Mass and initial separation have values, which are appropriate for the problem of explanation of the LIGO/Virgo events by coalescing binary PBHs. We consider both radiation dominated and CDM dominated stages of the evolution using numerical and semi-analytical means. We show that at the end time of our numerical simulations binary's semimajor axis decreases by approximately one hundred times, while its angular momentum decreases by ten times, in comparison to the standard values, which do not take into account effects associated with CDM clustering. We check that our conclusions are hardly affected by numerical artefacts. We estimate the merger rate of binary PBHs due to emission of gravitational wave at the present time both in the standard case when the effects associated with clustering are neglected and in the case when they are taken into account and show, that these effects could increase the merger rate at least by $6-8$ times in comparison to the standard estimate. This, in turn, means, that a mass fraction of PBHs, $f$, should be smaller than it was assumed before. 
\end{abstract}

\maketitle


\section{Introduction}
Primordial black holes (PBHs)
attract attention of many researchers for more than five decades. Depending on their mass they  can serve as a candidate for dark matter 
(see, e.g. \cite{dm0, dm1, dm2, dm3, dm4, dm5, dm6}), provide seeds for early formation of  supermassive black holes (see, e.g. \cite{smbh_seed1, smbh_seed2, polnarev85}), take an active part in the processes occuring in the Early Universe, see e.g. \cite{Carr_20_candi, Carr_21_candi} for a review and further references.
The interest to PBHs has been reinvigorated recently after detection of gravitational waves by the Advanced Laser Interferometer Gravitational-Wave Observatory (LIGO) \cite{ligo1, ligo2, ligo_merge, ligo3, ligo5, ligo4, ligo150}\footnote{LIGO/Virgo Public Alerts from the O3/2019 observational run can be found at https://gracedb.ligo.org/superevents/public/O3/}. Most of the mergers are associated with binary black holes with masses of order of a few tens of solar masses. It was suggested that  the binary black holes involved in these events may be of primordial origin, see e.g. \cite{Kovetz_17, Sasaki_16, Blinnikov_16,Mukherjee21}\footnote{Note, however, that some of the LIGO/Virgo events may have another origin, see e.g.  \cite{astro_bh}. It is important to stress that even if all LIGO/Virgo events will be
shown to be due to
mergers of black holes of stellar origin, if would give a very non-trivial constraint on the abundance of PBHs of stellar masses
in the Universe provided there is a reliable way of an estimate of the formation of PBHs pairs with orbital parameters appropriate for the LIGO/Virgo events for a given abundance of PBH. The exact possible mass fraction of PBHs in dark matter (DM) in LIGO/Virgo mass range is still a subject of large theoretical uncertainties, see, e.g. \cite{constr_war1, constr_war2}.}.

Therefore, a certain scenario of the formation of binary PBHs must be specified
in order to establish the connection between the merger rate needed to explain the LIGO/Virgo events by mergers of binary PBHs  and properties of PBH population, most importantly, their typical masses $M$  and mass fraction of PBHs, $f$, of total density of cold dark matter.

Perhaps, the most popular scenario of a kind  was put forward by Nakamura et al. in 1997 \cite{Nakamura_97}
in a different context. It is based on the idea  that if a pair of PBHs has an initial comoving distance between their members, $d_0$, smaller than an average comoving distance between PBHs due to statistical fluctuations, 
it might get bound at times comparable with
the time of equipartition between mass densities of cold dark matter (CDM) and
radiation, $t_{eq}$. This scenario was developed in the considered context by Sasaki et al. \cite{Sasaki_16} and Ali-Haïmoud et al. \cite{AKK} and discussed in some detail below, in the
Section \ref{sec:basic} \footnote{Among the alternative scenarios let us mention the scenario of the formation of the bound pairs in galactic cusps at relatively late time, see e.g. \cite{Bird_16, Fakhry_21}. Note, however,
that this scenario seems to require a large mass fraction
of PBHs $f\sim 1$, which may be refuted by the constraints 
on $f$ in this mass range, see e.g \cite{Carr_21_constr}.}.

The authors cited above neglected the evolution of the bound pairs due to gravitational interactions with the rest of dark matter. On the other hand, from their results it follows that, for typical values of $M$ and $f$, mass 
of CDM particles enclosed within $d_0$ of 
PBH pairs producing the observed merger rate  is comparable to $M$. In such a situation dynamical gravitational  interaction between the pair and CDM could
influence orbital parameters of the bound pair, and, in 
turn, estimates of the merger rate.

Perhaps even more important effect is the formation of 
halos both around isolated PBHs and, formally, around isolated bound pairs, immersed in an initially spatially uniform 
distribution of the ordinary CDM particles, see e.g. \cite{MOR_2007, Eroshenko_16} and references therein. 
The effect is more pronounced at relatively 
late times $t > t_{eq}$, when it may be shown that halo's
mass grows approximately proportional to the scale factor, $s$, and 
density distribution of CDM particles has a power law cusp
centered at the center of gravity. The presence of such cusps could clearly enhance the interaction between the pair
and CDM particles.

An attempt to answer the question to what extent the interaction between the pair and CDM particles could influence the merger rate was undertaken by Kavanagh et al. \cite{kavanagh} in 2018. The
authors came to the conclusion that this effect plays a minor role in estimates of the merger rates. 
However, their problem setting was rather unrealistic. They considered, by numerical means, two
bound massive particles representing PBHs placed at distances larger that periastron distance of a 
fiducial Kepler orbit and surrounded by two separate
halos of lighter particles with prescribed halo's masses and
profiles. The interaction between the pair and the particles
occurred mainly at periastron resulting in destruction of 
the halos and reduction of the pair semimajor axis by a factor of few and an insignificant change of orbital angular momentum. Neither the effects determined by cosmological expansion nor the effect of the formation and growth of the common halo around the pair were taken into account.  We are going to show below that taking halo growth into account is crucial in determining the evolution of pair parameters.

In this Paper we revisit the problem of interaction between a binary PBH and clustering 'ordinary' CDM in an idealized, but  self-consistent way. We consider initially unbound pairs of BHs following cosmological expansion at times $t\ll t_{eq}$.
These pairs get bound with time, forming a binary PBH \footnote{By a binary PBH we understand later on a gravitationally bound pair of PBHs. We use both these terms interchangeably below.}, and start to interact with the ordinary dark matter. We follow 
the evolution of the system, containing a PBH pair, dark matter and radiation, well into the matter dominated stage corresponding 
to times $t \gg t_{eq}$, using numerical and semi-analytical means. Typically, our simulations are stopped when the scale factor is ten times larger than its value at the equipartition time. However,
when we consider runs with pairs having larger initial separations, we finish our simulations earlier, 
since we do not find a significant evolution of the quantities of interest at late times in this case. During this  
evolution,  both binary's semimajor axis and its angular momentum get smaller than the values predicted by the purely analytical approaches due to interaction with CDM clustering around the pair. 
We find these values at the end of calculation and use them to make an estimate of the merger rate, which
is then compared with an analytical prediction. 

In our 
numerical work below interaction of the pair with other PBHs is neglected, as well as clustering of radiation and the standard  cosmological perturbations of CDM and radiation. Note, however, that we take into account interaction with 
other PBHs and cosmological perturbations in our estimates of the merger rate using the results of \cite{AKK}. This is needed because
our numerical scheme prohibits very small values of angular momentum that are predicted for the binary PBHs, which may potentially contribute
to the event rate. We check that the values of angular momentum obtained as a result of the numerical evolution are proportional
to the initial ones, calculate the ratios of these values and use analytical estimates for initial values of angular momentum to
calculate the final values expected for 'realistic' binary PBHs. In what follows we assume  that the mass fraction of PBHs is much smaller that that of the ordinary CDM, $f\ll 1$, and consider the mass densities of radiation and dark matter appropriate for the standard $\Lambda$CDM cosmology. We also assume that all PBHs have the same mass $M$. In this setting the problem is fully characterized by $M$ and the initial comoving distance between the members of the pair, $d_0$. We set $M=50M_{\odot}$ in 
all of our numerical runs and consider $d_0=0.5$, $0.95$, $1.5$ and $2$ in units of $\bar R=\sqrt[3]{\frac{3M}{4\pi \rho_{eq}}}$ (where $\rho_{eq}$ is the matter density at the matter-radiation equipartition), noting that our results can be extended to other values 
of mass by an appropriate redefinition of $d_0$.

Although we neglect any perturbations of the system other than those induced by the pair itself, there is an inevitable numerical 
noise. We check that this noise does not influence significantly our results by performing runs with different numbers of particles 
representing the ordinary CDM and considering runs with a single PBH with mass $2M=100M_{\odot}$. Only those results which are
stable with respect to a change of the number of  particles  are taken into account.

We find that the final binary angular momentum gets  
typically ten times smaller than its initial value, while the semimajor axis typically
decreases by $10^2$ times in comparison with the analytical results. These changes are apparently caused by two effects. At first,
a strong decrease of both quantities is observed when the members of the pair approach periastron at the first time. This may be due to expulsion of a significant mass of CDM particles from its vicinity during this time. Secondly, 
there is a secular change of these quantities 
afterwards. We discuss below that it may be due to the effect of
binary's 'hardening' due to gravitational interaction with CDM particles with sufficiently small angular momenta, which are 
constantly supplied to the halo, and, then, to the vicinity of binary PBH, from the ambient medium in course of the halo's growth.
While the former effect is more significant for the runs with larger values of $d_0$, the latter one is more pronounced for the
smaller values. We propose a quantitative semi-analytical model of the hardening process.

Depending on a value of $f$, our estimate of the merger rate, which takes into account these effects gives values $\sim 6-8$ times larger, than a value, which is obtained neglecting them. 
This, in turn, means, that in order to satisfy the observational limit 
on the merger rate, a value of $f$ should be a few times smaller than the standard estimate. We stress that these results should be
treated as a preliminary ones. Since the hardening proceeds until the end of our simulation, it could be effective at later times. 
This could lead to even larger values of the merger rate. On the other hand, the unaccounted effects of perturbations of orbits of CDM particles could influence our results in the opposite way due to the presence of other PBHs and cosmological perturbations.

The structure of the Paper is as follows. In Section \ref{sec:basic} we introduce our basic definitions, relations, and results of the solution of an equations describing dynamics of the pair in the absence of the effects of clustering CDM. Section \ref{sec:num} is devoted to the numerical simulations, while in Section \ref{implications} we make estimates of the merger rate. We conclude and 
finally discuss our results in Section \ref{disc}.

\section{\label{sec:basic}Basic definitions and relations}
\subsection{Notations and conventions}

In what follows we neglect the influence of 
primordial black holes (PBHs) on the overall expansion of the Universe, assuming that it mainly consist of radiation and dust-like cold dark matter (CDM) particles.  We are going to consider processes occuring at epochs before and after equipartition time, $t_{eq}\approx  23000yr$, defined by the usual condition that mass densities of matter and radiation are equal to each other at this time, and we denote these mass densities as $\rho_{eq}$. The scale factor, $s$, is assumed to be equal to one when $t=t_{eq}$. From the first Friedman equation we easily find a relation between $t_{eq}$ and $\rho_{eq}$
\begin{equation}
G\rho_{eq}t_{eq}^2=\frac{3\alpha^2}{ 16\pi}, \quad \alpha=\frac{2}{3}\sqrt {2}(2-\sqrt{2})\approx 0.552,
\label{e1}
\end{equation}
where $G$ is gravitational constant. Also, by integrating this equation we obtain
\begin{equation}
\sqrt{(1+s)}(s-2)+2=(2-\sqrt{2})\tilde t, \quad \tilde t=t/t_{eq},
\label{e2}
\end{equation}
as well as useful relations between $s$ and its first and second derivatives with respect to the time $\tilde t$
\begin{equation}
\dot s = \frac{2}{3}(2-\sqrt{2})({\frac{\sqrt{1+s}}{s}}), \quad \ddot s =-\frac{4}{9}{\frac{(2-\sqrt{2})^2}{s^3}}(1+{\frac{s}{2}}),
\label{e3}
\end{equation} 
 where dot stands for the derivative with respect to $\tilde t$. In the asymptotic limits $\tilde t \rightarrow 0$ and 
$\tilde t \rightarrow \infty$ from (\ref{e2}) it follows that $s\approx {({\frac{4}{3}}(2-\sqrt{2})\tilde t)}^{1/2}$ and
$s\approx {((2-\sqrt{2})\tilde t)}^{2/3}$, respectively.

In our analytical and numerical work we formally assume that a mass fraction of primordial black holes is $f<1$ and  all black holes have the same masses $M=50 M_\odot$. Our main results can be extended to other values of mass, therefore, we use dimensionless mass $M_{*} = \frac{M}{50 M_\odot}$ in our expressions below as a parameter. A physical distance between two 
particular black holes is denoted by $R$, while the corresponding comoving distance, $R/s$, is $D$. It is convenient to introduce an average
distance $\bar R=\sqrt[3]{\frac{3M}{4\pi \rho_{eq}}}$ 
and the associated dynamical time $t_d=\sqrt{\frac{{\bar R}^3}{GM}}$.
From eq. (\ref{e1}) we have 
\begin{equation}
t_d={\frac{2t_{eq}}{\alpha}}\approx 3.62t_{eq}.
\label{e4}
\end{equation}

Dimensionless physical and comoving distances are, respectively, $r=R/\bar R$ and $d=D/\bar R$. It turns out that the evolution of a separated 
binary black hole depends significantly on initial separation between binary components, 
$d_{0}=d(\tilde t\rightarrow 0)$. This quantity will
be used to characterize different initial conditions for our numerical simulations. 

\subsection{An equation describing the evolution of semimajor axis without the effect of CDM clusterization}
\label{equation}

When clusterization of CDM is not taken into account, one can use a simple ordinary differential equation to describe the evolution
of $R$, its time derivative, and, accordingly, binary's semimajor axis $a={\frac{GM}{|E|}}$, where $E$ is binary's binding energy per unit of mass\footnote{Note that the energy is not conserved initially, when tidal forces determined by expansion of the Universe dominate over gravity of the binary. Nonetheless, we use the usual Keplerian definitions of binding energy
and semimajor axis at all times.}, see e.g. \cite{1975MNRAS.173..729H}. In what follows we neglect orbital angular momentum of the binary,
formally assuming that its eccentricity equals to one, see e.g. \cite{Nakamura_97, Sasaki_16, AKK} for justification of this assumption. In this case the dynamical equation for the evolution of $R$ has
the form
\begin{equation}
\ddot R -{\frac{\ddot s}{s}}R=-{\frac{2GMt_{eq}^2}{R^2}},
\label{e5}
\end{equation}
where the second term on l.h.s describes cosmological tidal forces, its explicit form follows from (\ref{e3}), the term
on r.h.s is the usual Newtonian gravitational force per unit of mass, and we remind that dot stands for differentiation
over $t/t_{eq}$\footnote{Note that one can also, in principle, include the dynamic friction term in this equation. However, it can be 
shown to be small during the radiation dominated stage due to the effect discussed in \cite{Shukhman_1982}.}. We introduce $r=R/\bar R$, use the definition of $\bar R$ and eq. (\ref{e1}) to bring (\ref{e5}) to 
the form
\begin{equation}
\ddot r -{\frac{\ddot s}{s}}r=-{\frac{\alpha^2}{2 r^2}},
\label{e6}
\end{equation} 
which should be solved subject to the condition that $r \rightarrow s(\tilde t)d$ when $\tilde t \rightarrow 0$. Using
the definitions of binding energy and semimajor axis as well as (\ref{e1}) we can easily express them
in terms of the dimensionless distance $r$ and its time derivative 
\begin{equation}
E=\epsilon {\frac{{\bar R}^2}{t_{eq}}}, \quad a={\frac{\alpha^2}{4\epsilon}}\bar R, \quad \epsilon={\frac{\alpha^2}{2r}}-{\frac{{\dot r}^2}{2}}. 
\label{e7}
\end{equation}

In general, equation (\ref{e5}) should be solved numerically, and its solutions could be parametrized by values of 
the initial comoving distance, $d_0$. These solutions have the following qualitative behaviour. Initially, 
at sufficiently small values of $\tilde t$, the pair is unbound and the distance follows the Hubble expansion,
$r\approx s(\tilde t) d$.  However, the relative  importance 
of gravitational attraction 
of the companion black hole grows with time, and, at a certain time $t_{bind}$ defined by the condition
that $E(t_{bound})=0$, the pair gets bound, with its binding energy larger than zero. During the following stage
the semimajor axis monotonically decreases up to some approximately constant value $a_{fin}$. At later 
times, in the absence of interaction with accreting dark matter, $a_{fin}$ is assumed to be changed only by emission
of gravitational waves and we are going to compare our numerical results on the evolution of $a$ due to interaction
with CDM with $a_{fin}$. The stage of a 
significant decrease of $a$ terminates when $r$ becomes formally 
equal to zero at first time, and we denote this moment of time as $t_{fin}$. 

When $\tilde t$ is small it is easy to see that ${\frac{\ddot s}{s}}\approx {\frac{1}{4\tilde t}}$ on r.h.s. of (\ref{e5}). In this
case (\ref{e5}) is invariant with respect to a linear change of variables $r \rightarrow \beta r_n$ and  
$\tilde t \rightarrow \beta r_n$ provided that $\gamma^2=\beta^3$. The remaining freedom in choosing $\beta $ and $\gamma$ can be 
used to make the asymptotic behaviour of $r_n(\tilde t_n \rightarrow 0) \propto {\tilde t}^{1/2}d_0$ 
independent of $d_0$. It turns out that in 
this case one have to use $\beta \propto d_0^4$ and $\gamma \propto d_0^6$. After such a choice 
of $\beta$ and $\gamma$ is made, being considered in terms of the new variables 
$r_n$ and $\tilde t_n$ both equation (\ref{e5}) and the initial condition to it do not depend on any parameters, and, therefore,
the solution to this equation is uniquely fixed. The scaling  of all quantities characterising the solution expressed 
in terms of the old
variables, such as $s_{fin}$, $s_{bind}$ and $a_{fin}$ with $d_0$, should follow from the scaling of $\beta$ and $\gamma$. The scaling 
of $a_{fin}$ expressed in units of $\bar R$ should be the same as the distance $r$, and, accordingly,  $a_{fin}/\bar R\propto d_0^4$ 
while the scalings of $s_{fin}$ and $s_{bind}$ should be the same as 
the scaling of the scale factor $\propto \sqrt{\tilde t}\propto d_0^3$. The coefficients of proportionality in front of these scaling relations can be chosen by comparing them to the results of a numerical solution of (\ref{e5}). In this way we obtain that
\begin{equation}
a_{fin} \approx 0.1d_0^4\bar R, \quad s_{fin}\approx 0.56 d_0^3, \quad s_{bind}\approx 0.17d_0^3  
\label{e8}
\end{equation}
when $d_0$ is sufficiently small (see, also e.g. \cite{AKK}).      

\begin{figure*}
\includegraphics[width=0.6\textwidth]{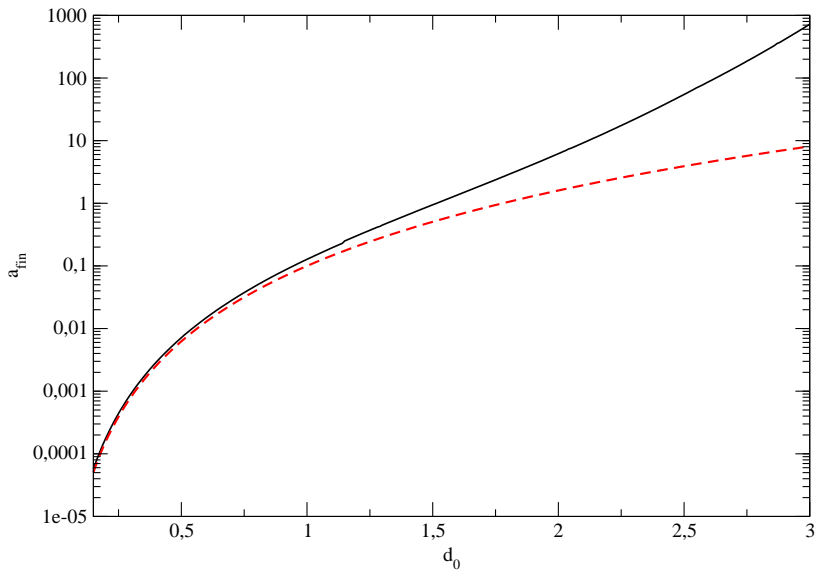}\vskip10pt
\caption{The dependency of the semimajor axis $a_{fin}$ expressed in units of $\bar R$ on $d_0$, shown as the solid line as
well as its approximate value valid at small $d_0$ and given by the first expression in (\ref{e8}) given as the dashed line.}
\label{fig1}
\end{figure*} 

\begin{figure*}
\includegraphics[width=0.6\textwidth]{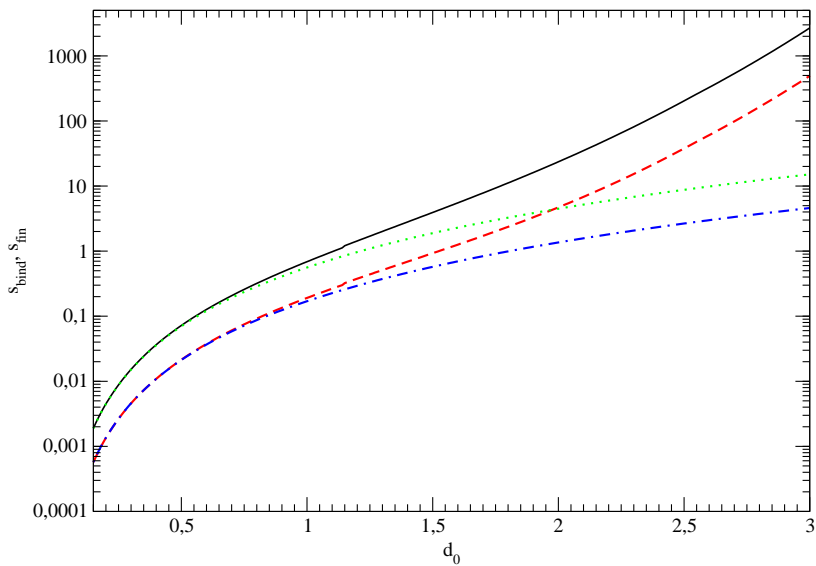}
\caption{The dependencies of $s_{bound}$ and $s_{fin}$ on the dimensionless initial comoving distance $d_0$ shown
as the 
solid and dashed lines, respectively. Additionally, we show as the dotted and dot-dashed lines $s_{fin}$ and $s_{bind}$ given 
by (\ref{e8}).}
\label{fig2}
\end{figure*}          
\bigskip
We plot $a_{fin}$ as a function of $d_0$
in Fig. \ref{fig1} and $s_{bound}=s(t=t_{bound})$ and  $s_{fin}=s(t=t_{fin})$ in Fig \ref{fig2} together with the corresponding
approximate relations (\ref{e8}). One can see from this Fig. that the relations (\ref{e8}) are indeed quite close to the numerical
solutions at $d_0 < 1$. Note that at large values of $d_0$ the effect of dark matter clusterization is very important and must be 
necessarily taken into account. 

\section{\label{sec:num}Numerical investigation of the evolution of PBH pair and surrounding CDM halo}
\subsection{\label{sec:ic}Simulation code and initial conditions}
In order to simulate an interaction of
PBHs and surrounding dark matter we use a direct summation N-body 4-th order Hermite code \texttt{ph4} which is a part of Amuse suite \cite{amuse1,amuse2}. This code is MPI-parallel and we made it also OpenMP parallel. Since we simulate systems at times close to the matter-radiation equality, we implemented radiation as a background. Our system also follows Hubble expansion at the beginning. In cosmological N-body codes this is usually implemented by solving particle motion equations in the comoving coordinates and using periodic boundary conditions. Since we are more interested in the behaviour of a gravitationally bound part of the system, we use integration in physical coordinates. Periodic boundary conditions will introduce a distortion to the simulation of a spherically symmetric system, so we use vacuum boundary conditions and our system is a sphere with finite radius, large enough to avoid effects of density discontinuity at the sphere boundary on the central part of the system. The sphere is centered at the origin of coordinates. Thus, in order to implement the cosmological background in the simulations, we introduce a radial acceleration, acting on particle at a position $\mathbf{r}$:

\begin{equation}
    \ddot{\mathbf{r}} = - {\frac{\alpha^2}{2}} s^{-4} t_{eq}^{-2} \mathbf{r}.
\end{equation}

We use a cubic grid for seeding the sphere with DM particles. Particles are assigned velocities according to the Hubble law. In case of a single black hole, it is placed close to the center with a tiny offset $<10^{-3}$ of the interparticle distance. If the BH is placed at (0, 0, 0) exactly, the trajectories of the closest particles pass directly through the BH and the timesteps of the simulation become very tiny, making it impossible to run. We do not take into account the absorption of DM particles by BHs in our simulations.

In case of a pair of BHs, we place them at comoving separation between BHs $d_0$, and assign tangential velocities assuming they are at apocenters of pure Keplerian orbits with eccentricity $e=0.8$.

We have made several test runs to check the simulation parameters and a series of production runs, which are summarized in a Table \ref{tab:sims}.

\begin{table*}[]
    \centering
    \begin{ruledtabular}
    \begin{tabular}{ccccccccc}
        Simulation &        $d_0$ & $\varepsilon$ & $s_{start}$ & $s_{end}$ & $N$ & $r_{max}$ & $m_p$ \\ \hline
        1BH\_50k\_eps0 &      0 & $4.3\times 10^{-6}$ & 0.1 & 16.2 & 51105 & 3.125 & $3\times10^{-4}$ \\
        1BH\_50k\_eps1 &      0 & $8.5\times 10^{-5}$ & 0.1 & 3.46 &  51105 & 3.125 & $3\times10^{-4}$ \\
        1BH\_50k\_eps2 &      0 & $8.5\times 10^{-3}$ & 0.1 & 3.46 & 51105 & 3.125 & $3\times10^{-4}$ \\
        1BH\_50k\_eps3 &      0 & $2.5\times10^{-2}$ & 0.1 & 3.46 & 51105 & 3.125 & $3\times10^{-4}$ \\
        1BH\_50k\_eps4 &      0 & $5\times 10^{-2}$ & 0.1 & 3.46 & 51105 & 3.125 & $3\times10^{-4}$ \\
        1BH\_50k\_early &      0 & $4.3\times 10^{-6}$ & 0.001 & 4.15 & 51105 & 3.125 & $3\times10^{-4}$ \\
        1BH\_200k &      0 & $4.3\times 10^{-6}$ & 0.1 & 8.7 & 203966 & 3.125 & $7.7\times10^{-5}$ \\
        2BH\_5k\_d0.95 & 0.95 & $4.3\times 10^{-6}$ & 0.1 & 8.1 & 4947 & 3.125 & $3\times10^{-3}$ \\
        2BH\_50k\_d0.95 & 0.95 & $4.3\times 10^{-6}$ & 0.1 & 16.2 & 51106 & 3.125 & $3\times10^{-4}$ \\
        2BH\_50k\_d0.95\_eps1 & 0.95 & $8.5\times 10^{-5}$ & 0.1 & 8.1 & 51106 & 3.125 & $3\times10^{-4}$ \\
        2BH\_50k\_d0.95\_eps2 & 0.95 & $1.7\times 10^{-2}$ & 0.1 & 8.1 & 51106 & 3.125 & $3\times10^{-4}$ \\
        2BH\_200k\_d0.95 & 0.95 & $4.3\times 10^{-6}$ & 0.1 & 10.6 & 203967 & 3.125 & $7.7\times10^{-5}$ \\
        2BH\_50k\_d0.7 & 0.7 & $4.3\times 10^{-6}$ & 0.001 & 5.6 & 51106 & 3.125 & $3\times10^{-4}$ \\
        2BH\_5k\_d0.5 & 0.5 & $4.3\times 10^{-6}$ & 0.001 & 8.1 & 4947 & 3.125 & $3\times10^{-3}$ \\
        2BH\_50k\_d0.5 & 0.5 & $4.3\times 10^{-6}$ & 0.001 & 16.2 & 51106 & 3.125 & $3\times10^{-4}$ \\
        2BH\_300k\_d0.5\_large & 0.5 & $4.3\times 10^{-6}$ & 0.001 & 7.4 & 299821 & 3.75 & $9\times10^{-5}$ \\
        2BH\_200k\_d0.5\_high & 0.5 & $4.3\times 10^{-6}$ & 0.001 & 1.1 & 203967 & 2.5 & $3.9\times10^{-5}$ \\
        2BH\_d0.5\_small & 0.5 & $4.3\times 10^{-6}$ & 0.001 &  8.1 & 20674 & 1.25 & $5\times10^{-5}$ \\
        2BH\_200k\_d1.5 & 1.5 & $4.3\times 10^{-6}$ & 0.1 & 12.4 & 203967 & 3.75 & $1.3\times10^{-4}$ \\
        2BH\_30k\_d2.0 & 2.0 & $4.3\times 10^{-6}$ & 0.1 & 38.2 & 31105 & 5.0 & $2\times10^{-3}$ \\
    \end{tabular}
    \end{ruledtabular}
    \caption{List of simulations with their parameters and abbreviations. $d_0$ is the comoving initial distance between BHs, $\epsilon$ is the gravitational softening parameter, $s_{start}$ and $s_{end}$ are scale factors at the beginning and the end of the simulation. $N$ is the total number of particles, including BHs, $r_{max}$ is the comoving radius of the Hubble sphere, $m_p$ is the particle mass divided by $2M_{*}$.}
    \label{tab:sims}
\end{table*}

We took caution of choosing a proper value of gravitational softening length $\varepsilon$. On the one hand, the DM is a collisionless fluid, and it is desirable to have $\varepsilon$ larger than the mean interparticle distance. On the other hand, we are interested in the behaviour of a pair of BHs, for which scattering and dynamical friction is important, so $\epsilon$ should not be too high. Having this in mind, we have conducted tests to check how our results depend on softening. We use the same value of softening length for both DM and BH particles. For a single BH, we have run 5 simulations with 50k particles and the values of $\varepsilon=4.3\times 10^{-6}$, $8.5\times 10^{-5}$, $8.5\times 10^{-3}$, $2.5\times10^{-2}$ and $5\times 10^{-2}$, respectively. The halo mass growth rate and density profile of these simulations is similar, while the anisotropy and distribution of angular momenta differ significantly. Both anisotropy and angular momentum are larger for simulations with higher $\varepsilon$ when we consider $\varepsilon=8.5\times 10^{-5}$, $8.5\times 10^{-3}$, $2.5\times10^{-2}$ and $5\times 10^{-2}$ respectively. The simulation with  $\varepsilon=4.3\times 10^{-6}$ demonstrates the same distributions of particle angular momentum and anisotropy as the simulation with $\varepsilon=8.5\times 10^{-5}$. As we discuss later on, angular momentum of infalling particles is a crucial parameter determining the evolution of a pair of BHs. From this analysis we conclude that values of $\varepsilon\leq 8.5\times 10^{-5}$ give stable results which are not affected by softening. For a pair of BHs we have conducted a similar investigation and found that the evolution of the pair semimajor axis is stable for $\varepsilon\leq 8.5\times 10^{-5}$. One should note that the mean interparticle distance at the halo center, estimated as $n^{-1/3}$, where $n$ is particle number density, always exceeds $10^{-3}$ for all our simulations.


Having a set of simulations with identical initial conditions and gravitational softening, but different number of DM particles, we investigate how well our results converge with the change of $N$, or particle mass $m_p$. As seen from Table \ref{tab:sims}, our simulations of a pair with $d_0=0.5$ cover almost 2 orders of magnitude in the particle mass, $m_p$.


\subsection{Properties of halos, formed around single or double BHs}
It is well known that in a homogeneous expanding Universe  a halo is formed around a single point mass (see, e.g., Gott 1975 \cite{1975ApJ...201..296G} \& Gunn 1977 \cite{1977ApJ...218..592G}). The halo boundary can be characterized by a turn around radius: at this radius a shell of matter turns around its radial velocity from Hubble expansion to infall. As has been shown in \cite{1985ApJS...58...39B}, the radius and mass of the halo grow with time as:
\begin{equation}
    r_\mathrm{ta} \propto s^{4/3},
\label{eq:rta}
\end{equation}
\begin{equation}
    M_\mathrm{ta} \propto s.
\label{eq:mta}
\end{equation}
We check that our simulations reproduce equations (\ref{eq:rta}) and (\ref{eq:mta}). At the same time we measure halo density profile from our numerical results. To do this, we construct a set of spherical shells centered at the black hole (or at the mass center of the pair, in the case of 2 BHs), each shell containing the same amount of particles $N_\mathrm{shell}=25$. The turn around shell is the furthest shell from the center which has mean radial velocity less than zero. The turn around radius is a radius at which the linearly interpolated mean shell velocity between the turn around shell and the next shell with positive radial velocity equals to zero. The turn around mass is the mass of all particles, except the black hole(s) inside the turn around radius.

\begin{figure*}
    \includegraphics[width=0.8\linewidth]{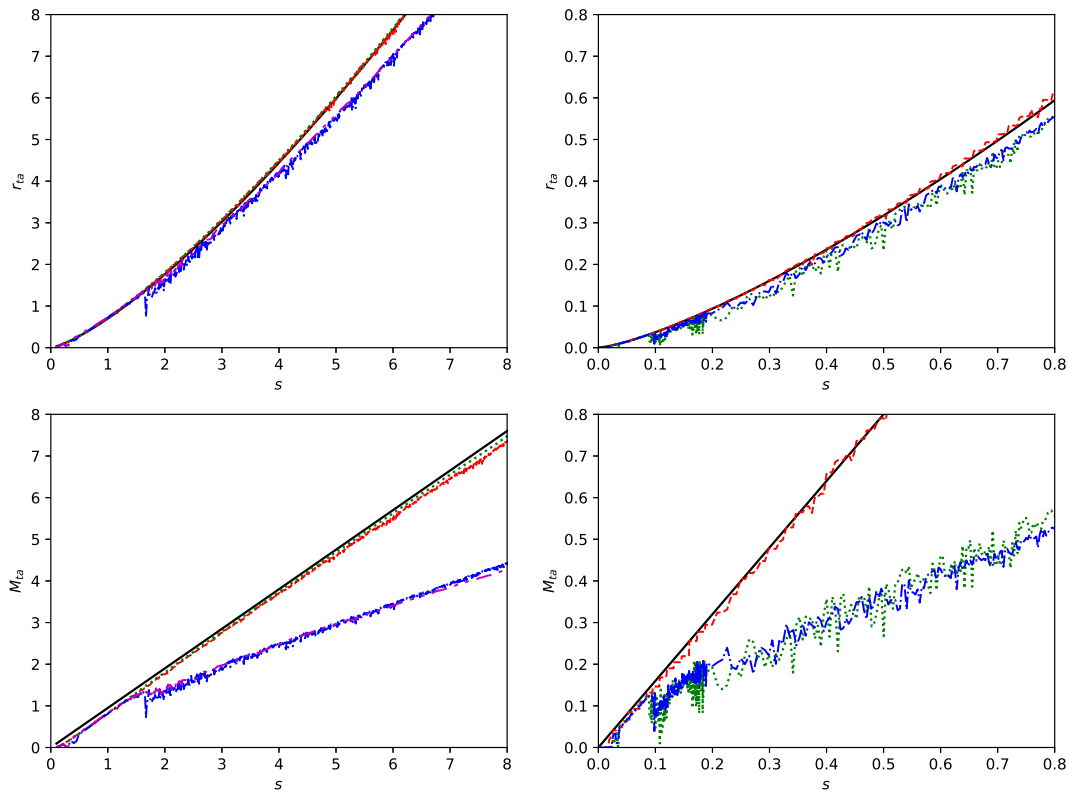}
    \caption{Left top and left bottom: evolution of the turn around radius and turn around mass in theory, $r_{ta}\propto s^{3/4}$, $M_{ta} \propto s$ (solid lines) and in simulations of single BH 1BH\_50k\_eps0 (dashed lines), 1BH\_200k (dotted lines), and the pairs with $d_0=0.95$, 2BH\_50k\_d0.95 (dot-dashed lines) and 2BH\_200k\_d0.95 (dot-dot-dashed lines). Right top and right bottom: the evolution of the turn around radius and turn around mass in theory (solid lines) and in simulations of the single BH 1BH\_50k\_early (dashed line) and the pairs with $d_0=0.5$, 2BH\_50k\_d0.5 (doted line), 2BH\_300k\_d0.5\_large (dot-dashed line).}
    \label{fig:rtamta}
\end{figure*}

In Fig. \ref{fig:rtamta} we show the evolution of $r_\mathrm{ta}$ and $M_\mathrm{ta}$ in several simulations. It is clearly seen that in simulations with a single BH the evolution of these characteristics very well reproduces the equations (\ref{eq:rta}) and (\ref{eq:mta}). Simulations with a pair also show a perfect match until some specific time which is close to the first pericenter passage of the pair. This passage happens at $s=1.15$ for simulations with $d_0=0.95$ and at $s=0.075$ for simulations with $d_0=0.5$. After that time, the halo radius becomes somewhat smaller, while the mass drops significantly. The turn around mass evolution of a halo around a pair of black holes is close to $M_\mathrm{ta}\propto s^{3/4}$ at late times. This demonstrates that the pair throws away a significant amount of matter.

\begin{figure}
    \includegraphics[width=0.95\linewidth]{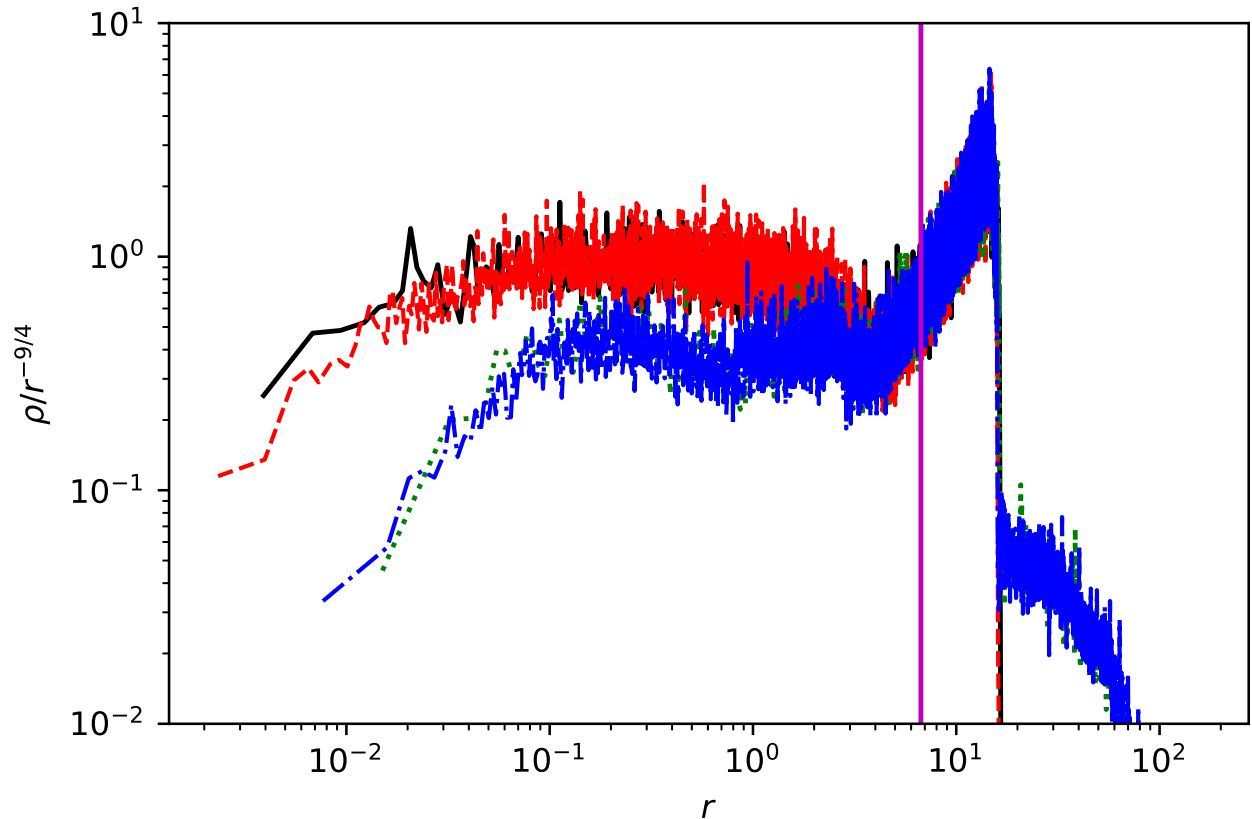}
    \caption{Dark matter density profiles of halos multiplied by $r^{9/4}$ at $s=5.6$ for simulations: 1BH\_50k\_eps0 solid line), 1BH\_200k (dashed line), 2BH\_50k\_d0.95 (dotted line) and 2BH\_200k\_d0.95 (dot-dashed line). Vertical black line indicates mean turn around radius of halos in these simulations at $s=5.6$.}
    \label{fig:densprof}
\end{figure}

Another test of simulations is the density profile of dark matter halos. Theory predicts the formation of a power law profile $\rho\propto r^{-9/4}$ \cite{1985ApJS...58...39B}. In Fig. \ref{fig:densprof} we demonstrate how our simulated halos follow this predicted law\footnote{Note that this result is different from that obtained in \cite{MOR_2007}, who claimed that $\rho\propto r^{-3}$}. It is seen that there is a good agreement, but the profile for a single black hole is flattening to $\rho\propto r^{-1.5}$ at $r<0.1$. For the pair of black holes this flattening is more pronounced, and also the whole profile has smaller values of density than the one corresponding to a single black hole at $r \lesssim 2 $, which is of the order of the radius of influence of the binary. The effect of flattening of the density profile is clearly determined by domination of 
gravitational field of the black hole or the binary at small distances. The evolution of halo mass and the dark matter profiles show a remarkable stability with the change of the numerical resolution, i.e. the particle mass $m_p$ over almost two orders of magnitude.

\begin{figure*}
    \includegraphics[width=0.8\linewidth]{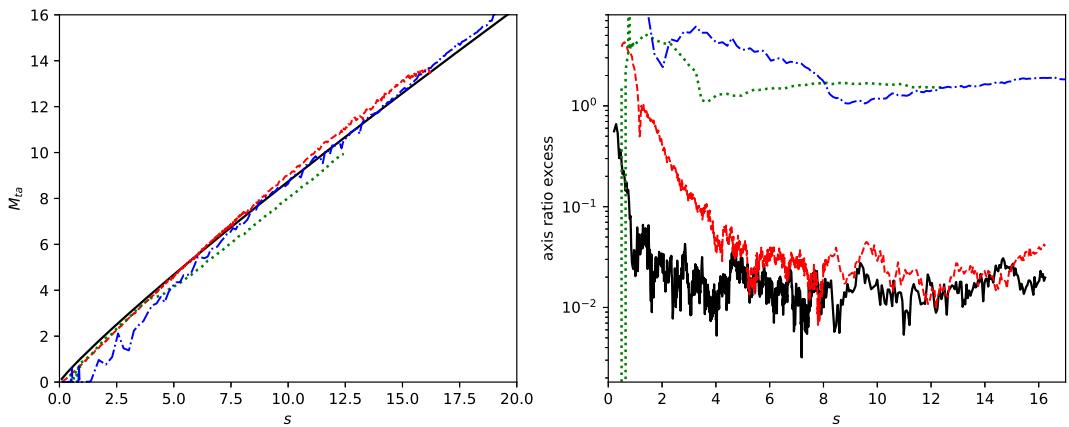}
    \caption{Evolution of accreted DM halos at late times. Left panel: mass of a halo inside a turn around radius, the approximation $M_{ta}\propto s^{0.9}$ (black solid line) and simulation results for 1 BH (1BH\_50k\_eps0, dashed line), and 2 BHs with $d_0=1.5$ (2BH\_85k\_d1.5, dotted line), and $d_0=2$ (2BH\_30k\_d2, dot-dashed line). Right panel: gyration tensor axis ratio excess at turn around radius for simulations with a single BH (1BH\_50k\_eps0, solid line), and for BH pairs with $d_0=0.95$ (2BH\_50k\_d0.95, dashed line), $d_0=1.5$ (2BH\_200k\_d1.5, dotted line) and $d_0=2.0$ (2BH\_30k\_d2, dot-dashed line). }
    \label{fig:late}
\end{figure*}

For the pairs of BHs with a large initial distances, $d_0=1.5$ and $d_0=2$, there is some notable differences in halo evolution. The two halos around each BH grow in isolation for a long period of time, gaining mass much larger than the mass of BHs they host. After these two halos merge, the resulting single halo shows growth of mass and radius quite similar with the case of a single BH and  it is well approximated as $M_{ta}\propto s^{0.9}$, see Fig. \ref{fig:late}. However, the halo itself becomes severely anisotropic. We demonstrate it by measuring the major to minor axis ratio of a reduced gyration tensor, defined as:
\begin{equation}
    G_{ij} = \frac{1}{N}\sum_{k=1}^N \frac{x^k_i x^k_j}{r^k r^k},
    \label{eq:gtensor}
\end{equation}
where $x^k_i$ is $i$-th coordinate of $k$-th particle. We compute $G_{ij}$ for all particles with $r<r_{ta}$, excluding the BH particles, then we find eigenvalues of $G_{ij}$. The axis ratio is defined as the square root of the ratio of maximal and minimal eigenvalues of $G_{ij}$. In Fig. \ref{fig:late} we subtract 1 from this axis ratio, i.e. value of 0 corresponds to a spherically symmetric distribution. The halos around BH pairs that have merged earlier are much more spherical, as seen from the evolution of the $G_{ij}$ axis ratio in 2BH\_50k\_d0.95 simulation in Fig. \ref{fig:late}. But the halo around a pair with $d_0=0.95$ is less spherical than a halo around a single BH at times $s<5$.

\subsection{Evolution of a BH pair}

We characterize the pair of BHs at every simulation snapshot by two main parameters: the semimajor axis of the orbit $\tilde a = a / \bar R$, where $a$ is defined in (\ref{e7}) and the dimensionless angular momentum $l$, which is computed as 
\begin{equation}
    l = v_t r,
\end{equation}
where $v_t$ is the relative tangential velocity of BHs (measured in units of $\bar R / t_{eq}$).

\begin{figure*}
    \includegraphics[width=0.8\linewidth]{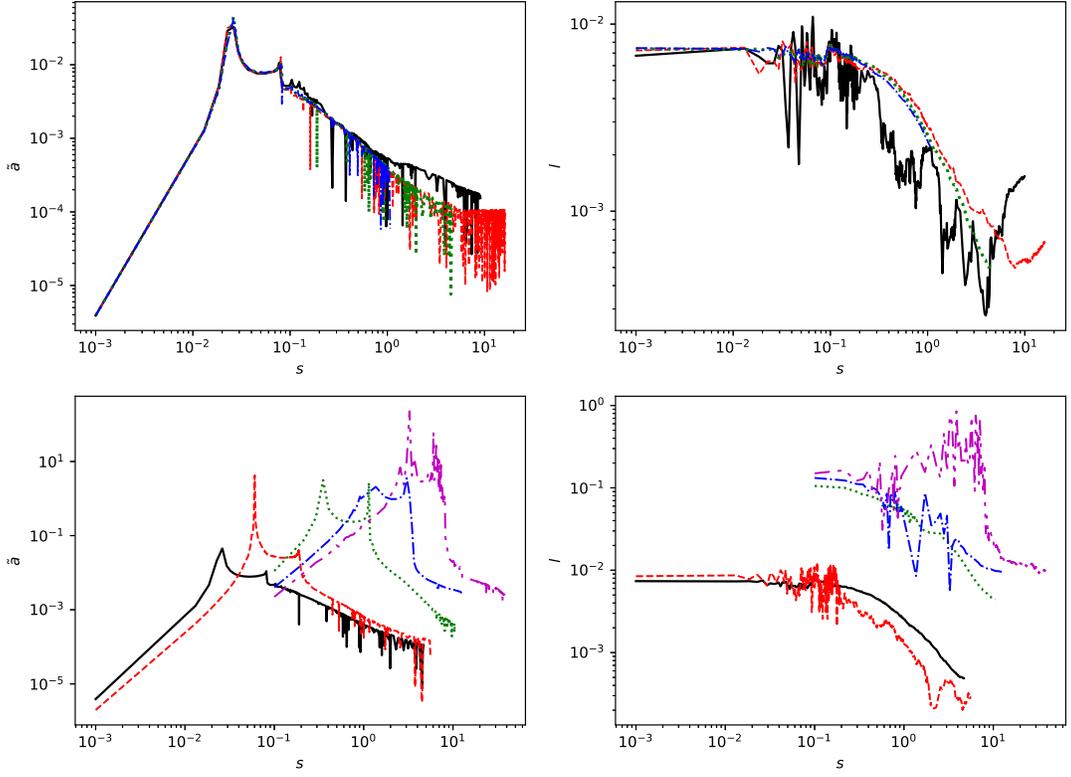}
    \caption{Top row: convergence of the evolution of dimensionless semimajor axis (left panel) and dimensionless angular momentum (right panel) in simulations with $d_0=0.5$ and $m_p=3\times 10^{-3}$ (solid line), $m_p=3\times10^{-4}$ (dashed line), $m_p=9\times10^{-5}$ (dotted line) and $m_p=3.9\times10^{-5}$ (dot-dashed line).
    Bottom row left panel: evolution of a semimajor axis of a BH pair in simulations with increasing initial comoving distance $d_0=0.5$ (solid line), $d_0=0.7$ (dashed line), $d_0=0.95$ (dotted line), $d_0=1.5$ (dot-dashed line), $d_0=2$ (dot-dot-dashed line). Bottom right panel: evolution of the dimensionless angular momentum $l$. The meaning of line styles and colors is the same as in the left panel.}
    \label{fig:simpair}
\end{figure*}

The results are shown in Fig. \ref{fig:simpair}.
The top row demonstrates the numerical convergence over almost two orders of magnitude in test particle mass in case of simulations with $d_0=0.5$. All simulations with the total number of particles from $5\times 10^3$ to $3\times 10^5$ show a steady decrease of $\tilde a$ after the first passage of pericenter, which happens at $s\approx 0.08$.
Note that this value is very close to the corresponding 
value $s_{fin}$ obtained from the solution of eq. (\ref{e6}), see Fig. \ref{fig2}. One should note that due to the decrease of $\tilde a$ with time, at some moment there is less than 1 DM particle between the BHs which affects the pair evolution. For the lowest resolution simulation with $m_p=3\times 10^{-3}$ this happens at $s=0.2$, while for simulations with $m_p=3\times10^{-4}$ and  $m_p=9\times10^{-5}$ this happens at $s=1$ and $s=2$, respectively. Our highest resolution simulation with $m_p=3.9\times10^{-5}$ is stopped before the regime with less than 1 particle between BHs is reached.

In the bottom left panel of Fig. \ref{fig:simpair} it is seen that simulations with $d_0<1.5$ show a steady decrease of $\tilde a$ after the first pericenter passage. Simulations with $d_0\geq1.5$ show a rapid drop of $\tilde a$ at the pericenter passage, determined by the merging of two massive halos.

The evolution of the dimensionless angular momentum, $l$, in Fig \ref{fig:simpair} shows two sets of lines with different initial $l$. One set corresponds to simulations which start at $s_\mathrm{start}=0.001$ and another one corresponds to $s_\mathrm{start}=0.1$. This is caused by the preparation of the initial conditions with fixed eccentricity described in Section \ref{sec:ic}. Our results show an approximately tenfold drop of angular momentum during the evolution of the system. 

At a first glance our numerical results for the evolution of $\tilde a$ and $l$ are at odds with the results of \cite{kavanagh}. To investigate the cause of the difference, we first run with our code a simulation with the initial conditions from \cite{kavanagh}, which has the PBH mass 30~M$_\odot$, the initial semimajor axis $a_i=0.01$~pc and eccentricity $e_i=0.995$. These initial conditions have only 6.2~M$_\odot$ mass in the dark matter, which is about 1/10 of the binary mass. We obtain results in a fine agreement with that shown in \cite{kavanagh}, so the difference with our simulations is not due to the fact that we use a different N-body code. 
We also run a simulation with the same initial conditions, but with the additional force caused by the radiation background. This gives almost no effect on the evolution of the binary.

\begin{figure*}
    \includegraphics[width=0.6\linewidth]{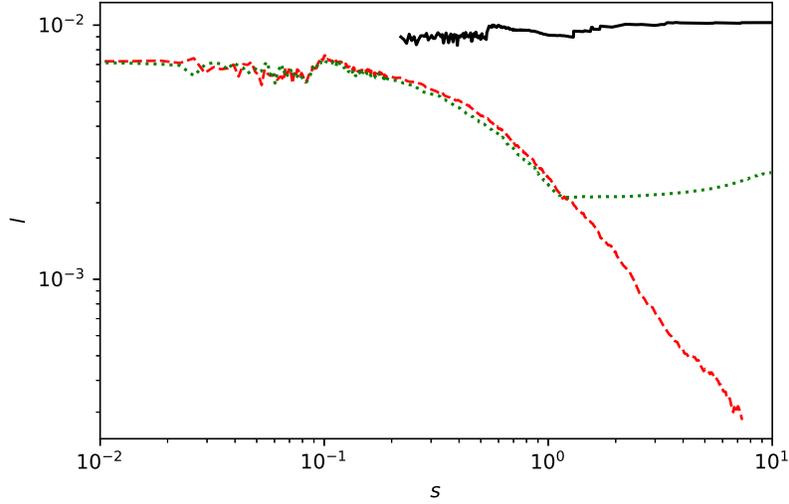}
    \caption{Evolution of the angular momentum in the simulation from the initial conditions from \cite{kavanagh} without cosmological expansion and DM accretion (solid line), in our simulation 2BH\_300k\_d0.5\_large with DM accretion during the full run (dashed line), and in a simulation where the accretion stops due to a smaller size of the Hubble sphere, 2BH\_d0.5\_small (dotted line).}
    \label{fig:lstop}
\end{figure*}

Next we check the influence of accretion of DM particles from the Hubble flow. In simulations of \cite{kavanagh} there is no such accretion, but according to a simple estimate, the halo around a binary should grow by ten times during the time period shown in Fig. 6 of \cite{kavanagh}. We run a test simulation with a Hubble sphere filled by DM particles, as described in Section \ref{sec:ic}, but the sphere radius is artificially chosen small enough to end the accretion in the middle of the simulation (this simulation is labelled as 2BH\_d0.5\_small in Table \ref{tab:sims}). We show in Fig. \ref{fig:lstop} that when the accretion stops, the angular momentum stops decreasing. The same happens for the semimajor axis. So, we conclude that the accretion of DM particles is the main factor responsible for the late evolution of the PBH pair, after the formation of the common halo. The infalling DM particles have near radial trajectories and thus persistently refill loss cone of the binary, see
also the following Section.

\subsection{A semi-analytic treatment of the late time evolution of semimajor axis and angular momentum}

\begin{figure*}
\vskip20pt
\includegraphics[width=0.6\textwidth]{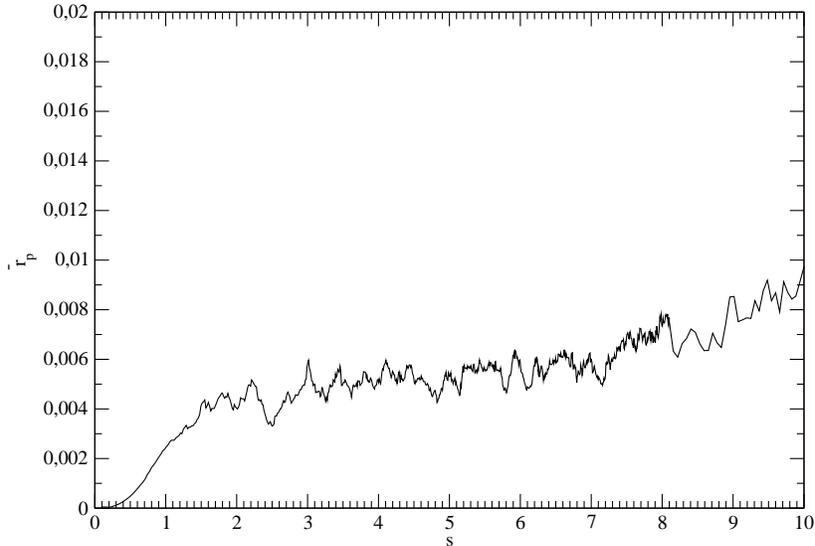}\vskip10pt
\caption{The dependency of the average periastron radius $\bar r_p$ on $s$, taken from the simulation $2BH\_200k\_d0.95$}
\label{rp}
\end{figure*}

We can see from Fig. \ref{fig:simpair} when $d < 0.95$ both semimajor axis and angular momentum experience a steady decrease. As discussed above, 
the most likely explanation for this behaviour is
due to 'hardening' of the binary due to its interaction with CDM particles having their periastron distances, $r_p$, order of or smaller than binary's semimajor axis $a$. 
As discussed in e.g. \cite{sesana}, the contribution of a given particle to the evolution of both quantities is determined by the ratio
$r_p/a$ (or, alternatively, the ratio of particle specific angular momentum to that of the binary), and only the particles with $r_p/a \sim 1$ can influence it significantly.
In principle, the origin of particle's angular momentum could be due to the tidal torque exerted by the binary on the particles. However, we checked that a value of angular
momentum expected from this effect is too small to explain what is observed in simulations.  A relatively large value of the angular momentum may be explained by torques 
acting on the particles from the side of non-spherical part of density distribution of the halos related to the observed
halo's anisotropy. The anisotropy itself may be originated from the action of the radial orbit instability (see e.g. \cite{2017MNRAS.470.2190P}) 
with initial perturbations provided by the action of binary's quadrupole gravitation field. In this picture the smaller is $d_0$ the smaller would
be the anisotropy and this is indeed observed in the simulations. An analytical calculation of the corresponding distribution of angular momentum is a non-trivial problem, which is left for a future work. In this paper we use an average value of $r_p$, $\bar r_p$, evaluated numerically at the 'influence radius' determined by the condition that halo's mass inside this radius is equal to that of the binary.  We show the dependency of $\bar r_p$ on $s$ in Fig. \ref{rp}, for the case $d_0=0.95$. 

Sesana et al. \cite{sesana} showed numerically that, a contribution of a particle with a given value of $r_p$ to the hardening rate of semimajor axis is determined 
by a function $C(x)$ of $x=\sqrt{r_p/a}$, while the same contribution to the evolution of angular momentum is given by a similar function $B(x)$. We used their numerical results
presented in their Figs 1 and 2, for an equal mass binary with eccentricity $e=0.9$ and fitted $C(x)$ by a simple function, $C(x)=2.647(1+x^2)^{-2}$, which is in a quite good
agreement with the numerical curve. $B(x)$ is a non-monotonic function of $x$, and, a simple fit of it does not appear to be possible. However, for our purposes we only 
need to know the integral $I(x)=\int_{0}^{x}dy yB(y)$, which can be fitted as $I(x)=2.86(1-\frac{1}{{(1+x^3)}^{3/5}})$ remarkably well. 

When $C(x)$ and $B(x)$ are known, it follows from the discussion in Sesana et al. \cite{sesana}
the time evolution of $a$ and $L$ are given by the simple expressions
\begin{equation}
\frac{\dot a}{a}=-\int^{\infty}_{0}C(x)d{\it P}_{r_{p}}\frac{\dot M}{M}, \quad \frac{\dot L}{L}=-\int^{\infty}_{0}B(x)d{\it P}_{r_{p}}\frac{\dot M}{2M},
\label{hard1}
\end{equation} 
where ${\it P}_{r_p}(r_{p})$ is a probability of finding the periaston radius with a value smaller than $r_p$. We model the corresponding probability density as a 
step function, $d{\it P}_{r_p}(r_{p})/dr_{p}=\frac{1}{2\bar r_p}\Theta (r_2-2\bar r_p)$, where the factor $\frac{1}{2\bar r_p}$ in the front of the step function
$\Theta (x)$  is chosen in such a way that the average value of $r_p$ coincides with $\bar r_p$. $\dot M$ in (\ref{hard1}) is the mass flow through the sphere of
influence of the pair (i.e. a sphere centered at the center of mass of the binary with the radius equal to the influence radius,
$r_{inf}\approx 2$, see the discussion above). We assume that it is approximately 
the same as the mass flow through the turn around radius, $\dot M \approx M\frac {ds}{dt}$. 

Combining all ingredients discussed above together, we can write down two evolution equations for $a$ and $L$:

\begin{eqnarray}
\frac{1}{a}\frac{d a}{ds}=-1.32\frac{a}{\bar r_p}(1-\frac{1}{1+\frac{2\bar r_p}{a}}), \quad \nonumber\\
\frac{1}{L}\frac{d L}{ds}=-1.43\frac{a}{\bar r_p}(1-\frac{1}{{({1+{(\frac{2\bar r_p}{a}})}^3)}^{3/5}}).
\label{hard2}
\end{eqnarray}

\begin{figure*}
\includegraphics[width=0.6\textwidth]{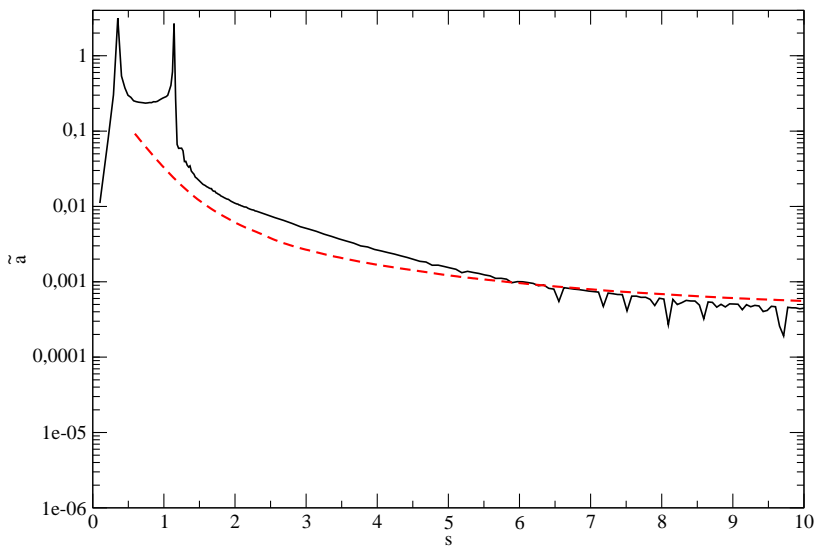}
\caption{The evolution of semimajor axis for the case $d_0=0.95$. Solid and dashed
curves represent the fully numerical result and the result of solution of (\ref{hard2}), respectively.}
\label{rt}
\end{figure*}

\begin{figure*}
\includegraphics[width=0.7\textwidth]{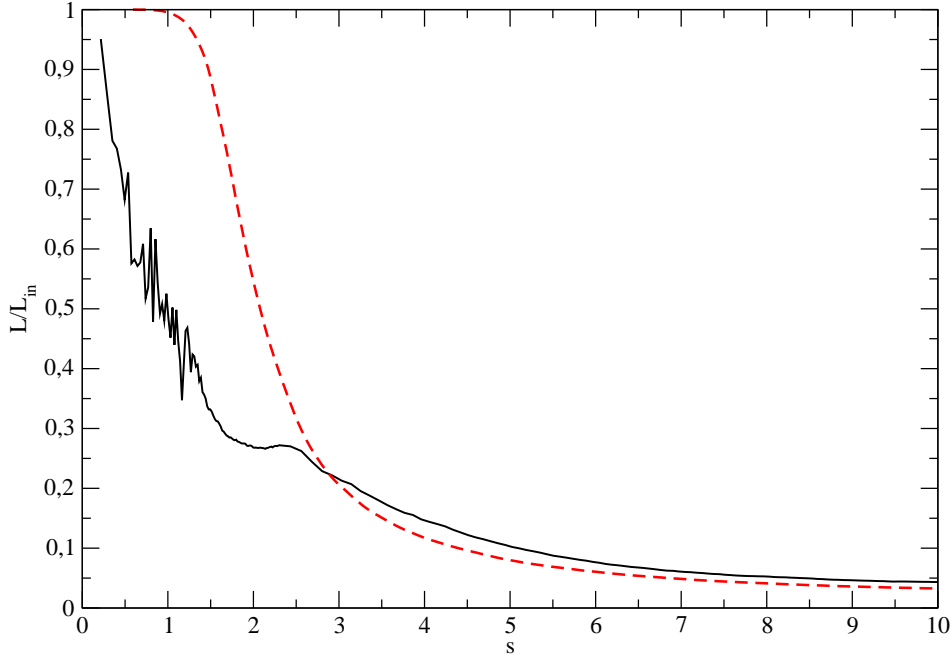}
\caption{The same as Fig. \ref{rt}, but for the evolution of angular momentum.}
\label{lt}
\end{figure*}

 As an initial condition we use values $a_{fin}$ and $s_{fin}$
obtained from the solution of eq. (\ref{e6}), $s_{fin}\approx 0.6$ and 
$a_{fin}(s_{fin})\approx 0.1$
Results of the solution of eqns (\ref{hard2}) are shown in Figs \ref{rt} and \ref{lt}. One can show from these Figs. that, indeed, at sufficiently large
values of $s$ the fully numerical results are quite close to the solutions
of (\ref{hard2}). 

\bigskip

\bigskip

\section{Astrophysical implications}
\label{implications}
In order to see to what extent the additional evolution of semimajor axis and angular momentum due to dynamic influence of CDM accretion 
would influence the results concerning the possibility of
explaining the LIGO events by PBHs, we need to estimate the formation rate of gravitational wave events, $R$ (or, the
merger rate),
both, for the case when the effects of CDM clustering are absent and in the case when they are present. In the former
case we closely follow  Ali-Haïmoud, Kovetz $\&$ Kamionkowski 2017 \cite{AKK} (hereafter, AKK), while in the latter case we
use the numerical results obtained above. For the comparison with observations we are going to consider only $M_{*}=1$, and
assume that when $M_* \sim 1$ observations constraint the rate  to be in the range 10 $Gpc^{-3}yr^{-1} < R <$ 100 $Gpc^{-3}yr^{-3}$, see e.g. \cite{AKK, ligo_merge}.

\subsection{Time scale of the evolution of semimajor axis} The formation rate $R$ is determined,
to a large extent, by a characteristic time scale of the evolution of $a$, $t_{GW}$, for a bound PBH pair having 
a large value of initial eccentricity $e\sim 1$
\begin{equation}
t_{GW}={\frac{3}{170}}{\frac{c^5}{{(GM)}^3}}a^4j^7,
\label{e9}
\end{equation} 
where $j=\sqrt{1-e^2}$, see e.g  \cite{AKK}. 

The pairs have a time to merge only when $t_{GW} < t_{H}$, where $t_{H}\approx 1.38\cdot 10^{10}yr$ is the present age of 
the Universe. We introduce $\tilde t_{GW}=t_{GW}/t_{H}$ and express it in terms of the quantities appropriate for the problem at hand. We obtain
\begin{equation}
\tilde t_{GW}=5.22\cdot 10^{19}j^{7}{\tilde a}^4M_{*}^{-5/3},  
\label{e12}
\end{equation}
where $\tilde a=a/\bar R$ and $M_{*}=M/50M_{\odot}$. From the condition $\tilde t_{GW} < 1$ we obtain
\begin{equation}
\tilde a < \tilde a_{crit}, \quad \tilde a_{crit}=1.17\cdot 10^{-5}j^{-7/4}M_{*}^{5/12}.  
\label{e13}
\end{equation}

In what follows we assume that both $\tilde a$ and $j$ 
(or, equivalently, the dimensionless angular momentum $l=\sqrt{\tilde a}j$) are certain functions of the initial comoving distance $d_0$, and a possible form of these functions follow from a particular scenario of the evolution of the pair before 
the process of its evolution due to emission of gravitational waves sets in. Note that it is different from the approach 
undertaken by AKK, who considered $j$ as a stochastic variable, but we are going to show that both approaches give quite similar
results in the scenario when the effects of accreting CDM are neglected. When these functions are specified the condition
$t=t_{GW}$ provide a certain value of $d_0$ (for given $f$ and $M_*$), $d_{*}(t)$, and only the pairs with $d < d_{*}(t)$ can merge before time $t$. When $t=t_{H}$ we set $d_{*}\equiv d_{*}(t_{H})$.

Assuming that spatial distribution of PBHs obey the Poisson statistics and introducing the variable $X=fd_0^3$ instead of $d_0$, one can easily show that probability density of distribution over a distance to the closest neighbor having some value of 
$X$, $\frac{d{\it P}}{dX}$ is $\frac{d{\it P}}{dX}=e^{-X}$. Accordingly, the probability to have a pair with $d_0 < d_*$, ${\it P}_*$, 
is ${\frac{1}{2}}X_{*}e^{-X_{*}}$, where the factor ${\frac{1}{2}}$ is inserted to avoid 
double counting of the components, and $X_*=fd_*^3$, see also AKK. Taking into account that we expect $X_{*} \ll 1$ we can set
${\it P}_*=X_*$.

In order to estimate the merger rate we need to know the present day number density, $n_{pbh}$. 
This follows from our definition of the PBH mass fraction $f$ and the average radius $\bar R$
\begin{equation}
n_{pbh}= \frac{3f}{4\pi {(s_{pt}\bar R)}^3}. 
\label{en1}
\end{equation}  
where $s_{PT}\approx 5.88\cdot 10^3$ is the present day value of the scale factor and we remind that we normalize it by the condition that it is equal to one at the equipartition time. A number density of PBHs, which have been merged before the
present time, $n_{merge}$, follows from (\ref{en1}) and the definition of our probability ${\it P}_*=X_*$
\begin{equation}
n_{merge}=\frac{3}{8\pi}\frac{fX(d_{*})}{{(s_{pt}\bar R)}^3}. 
\label{enn1}
\end{equation} 
We obtain the merger rate differentiating $X$ in (\ref{enn1}) and remembering that $X$ is a function of time due to the
condition $t=t_{GM}$ and our assumption that $\tilde a$ and $j$ entering $t_{GW}$ are functions of $d$. We have
\begin{widetext}
\begin{equation}
R= \frac{3}{8\pi}\frac{fX(d_{*})}{{(s_{pt}\bar R)}^3t}
(\frac{t_{GW}}{X_{*}}\frac{\partial X_{*}}{\partial t_{GW}})
\approx
2.86\cdot 10^{7}fX_*M_*^{-1}
(\frac{t_{GW}}{X_{*}}\frac{\partial X_{*}}{\partial t_{GW}})
Gpc^{-3}yr^{-1}. 
\label{en2}
\end{equation}
\end{widetext}

\subsection{The merger rate in case when CDM accretion is neglected} 
\label{standard}

When the influence of accreting CDM on the evolution of semimajor axis and angular momentum of the binary is neglected the problem can be analyzed with the help of solutions to equation (\ref{e5}) and calculation of some integrals related to those solutions, see e.g. AKK, \cite{AKK}. We closely follow AKK 
in this Paper for an estimate of the event rate in this case and use this estimate as a 'reference value' for the problem, which takes into account the effect of clusterization. 

When the accretion is neglected one can use $a_{fin}$ represented in Fig. \ref{fig1} as the corresponding value of semimajor axis entering (\ref{e12}).  In the same approach $j$ does change significantly at time $t > t_{eq}$ and its value is assumed to be determined by two factors
- 1) gravitational torques on the pair from the side of other black holes, and 2) gravitational torques determined by the cosmological density perturbations. In the first case a distribution over possible values $j$ is given by equation (19) of AKK, which can be used to
calculate a mean value of $j$. However, the corresponding integral is logarithmically divergent, which is related to the fact that nonphysical
values of $j > 1$ are allowed. We set, accordingly, the upper integration limit $j=1$ there, thus obtaining the average value
\begin{equation}
j_{bh}=0.5fd^3\Lambda, \quad \Lambda=\ln {\frac{4}{fd^3}},   
\label{e10}
\end{equation}
where we remind that $f$ is the ratio of PBH mass density to CDM mass density and it is assumed that $f \ll 1$. When $f \ge 10^{-3}$ $\Lambda \sim 10$. 
The average value of $j$ due to the influence of cosmological perturbations, $j_{cp}$ is given by eq. (21) of AKK. Being rewritten
with the help of our notations it has the form
\begin{equation}
j_{cp}=2.5\cdot 10^{-3}d^3,  
\label{e11}
\end{equation}
where we assume that $\sigma_{eq}=5\cdot 10^{-3}$ in eq. (21) of AKK. In order to find a typical value of $t_{GW}$ in eq. (\ref{e9}) we use $j=j_{bh}$ when $j_{bh} > j_{cp}$  and  $j=j_{cp}$ otherwise. 


When the result of solution of eq. (\ref{e5}) and $j$ given by the expressions above are substituted in (\ref{e13}) the condition $\tilde a_{crit}=\tilde a_{fin}$ can be used to express $d_{*}$ as a function of $f$, for a fixed value of $M_*$.    
\begin{figure*}
\includegraphics[width=0.7\textwidth]{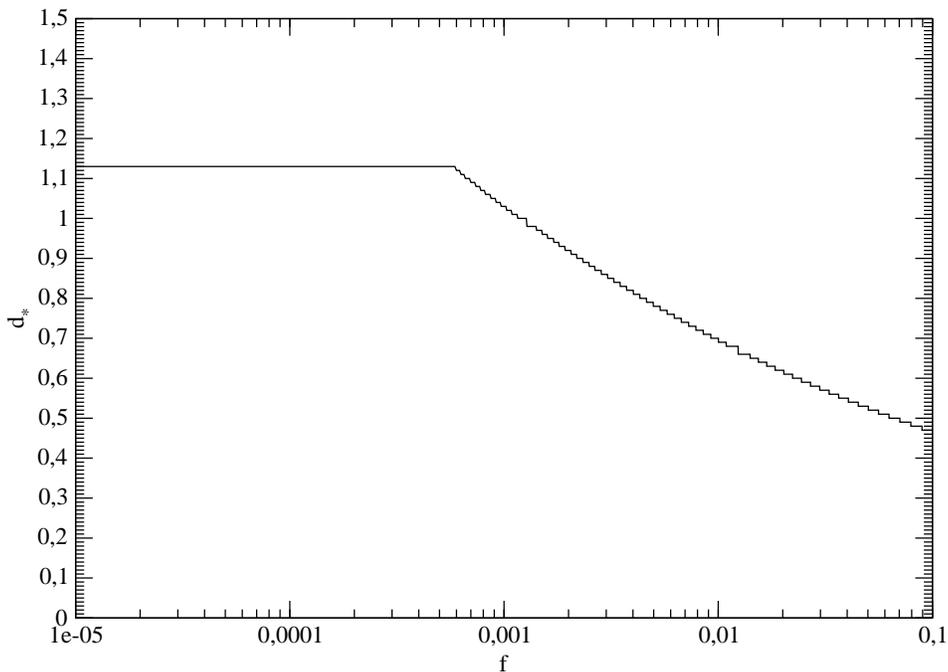}
\caption{The quantity $d_*$ as a function of $f$ for $M_*=1$.}
\label{fig3}
\end{figure*}          
\bigskip
We show this function in Fig. \ref{fig3} for the case $M_{*}=1$. Note that the saw-like behaviour of the curve is simply due to relatively small numbers of our grid points used to represent our quantity $d_{0}$ when solving (\ref{e5}). This is clearly nonphysical, but, since it does not influence any our results it is left unchanged. Also note that $d_{*}$ is constant at small values of $f$. This is because when
$f$ is small $j=j_{cp}$ 
which does not depend on $f$, see eq. (\ref{e11}) 

From eq.(\ref{e8}) it follows that $\tilde a_{fin}\propto d_*^4$ when $d_{*}$ is not too large, while from the discussion above
we see that we may approximate the dependence of $j$ on $d_8$ as $j\propto d^3$. Taking into account that $X_*=fd_*^3$ we
see from eq. (\ref{e9}) that $t_{GW}\propto X_*^{37/3}$ and, accordingly,
$\frac{t_{GW}}{X_{*}}\frac{\partial X_{*}}{\partial t_{GW}}=\frac{3}{37}$
in (\ref{en2}) when accretion is neglected.  
We use 
$\frac{t_{GW}}{X_{*}}\frac{\partial X_{*}}{\partial t_{GW}}=\frac{3}{37}$
in eq. (\ref{en2}), thus obtaining the event rate in the model case of the absence of accretion, which is shown in Fig. \ref{fig4} below together with an estimate of the event rate, which takes into account the effects
of CDM accretion. 
From this Fig. it is seen that, in the case when accretion is neglected, 
in order to have $R$ in the interval 10-100 $Gpc^{-3}yr^{-1}$ the mass fraction $f$ should be in the interval
$2.5\cdot 10^{-3}-1.2\cdot 10^{-2}$. It agrees perfectly with the corresponding result of AKK 
\footnote{Note that
the value of $M_*$ in their Fig. 5, which is the closest to our value $M_*=1$, is $M_*=3/5$. This shouldn't, however, influence the agreement in any 
significant way}, see their Fig. 5.
From Fig. \ref{fig3} it also follows that when $f$ is approximately in the range $10^{-3}-10^{-2}$, $d_*$ 
should approximately be in the range $0.5-1$. This justifies the range of $d_0$ chosen for our numerical simulations. 

\subsection{The merger rate in the case of CDM accretion} 

\subsubsection{Numerical evaluation of the quantities of interest and the corresponding fitting procedure}

In order to estimate the influence of the CDM accretion on the event rate we performed five simulations with $d_0=0.5$, $0.7$, 
$0.95$, $1.5$ and $2$. From the expression (\ref{e9}) it follows that $t_{GW}$ is determined by the dimensionless angular momentum
$l=\sqrt{\tilde a}j$ and $\tilde a$. Since the estimates (\ref{e10}) and (\ref{e11}) show that the pair is expected to be very
eccentric, 'realistic' value of initial eccentricity are numerically extremely expensive. Therefore, in order to estimate the evolution
of $l$ we use the following procedure: for each value of $d_0$ we calculate some initial value of $l$, $l_{in}$ during an initial stage
of numerical evolution and some final value $l_{fin}$ at some final time, $t_{fin}$. Then we calculate the ratio $\lambda =l_{fin}/l_{in}$
and multiply it by $\sqrt{\tilde a_{fin}}j$, where $\tilde a_{fin}$ is determined by the solution of equation (\ref{e6}) and $j$ is given
by  (\ref{e10}) or (\ref{e11}), as discussed above. The numerical value of $\tilde a$ used in (\ref{e9}), $\tilde a^{num}_{fin}$ is calculated at the same moment of time. This moment of time is taken to correspond to $s_{fin}\sim 10$ for the runs with $d_0 < 1$ and $s_{fin} \sim 5$ for the runs with  $d_0 > 1$, since in the latter case there is no significant evolution of $\lambda $ and $\tilde a$ further on. We have checked that numerical effects are not expected to influence our results both by performing runs with significantly different values of CDM particles and by comparing the results to the run corresponding to a single black hole. It is important to stress that the procedure
of evaluation of the final angular momentum explicitly implies the assumption that  torques acting on the pair are proportional to its
angular momentum. Although it is supported by theory when binary evolves in the hardening regime, see equation (\ref{hard1}), and finds
some evidences in our numerical simulations, since the runs corresponding to $d_0=0.5$ have much smaller initial values of angular momentum,
see Fig. \ref{fig:simpair}, a caution should be taken with respect to this point. We are going to check this assumption with special purpose
simulations in our future work.

\begin{figure*}
\includegraphics[width=0.7\textwidth]{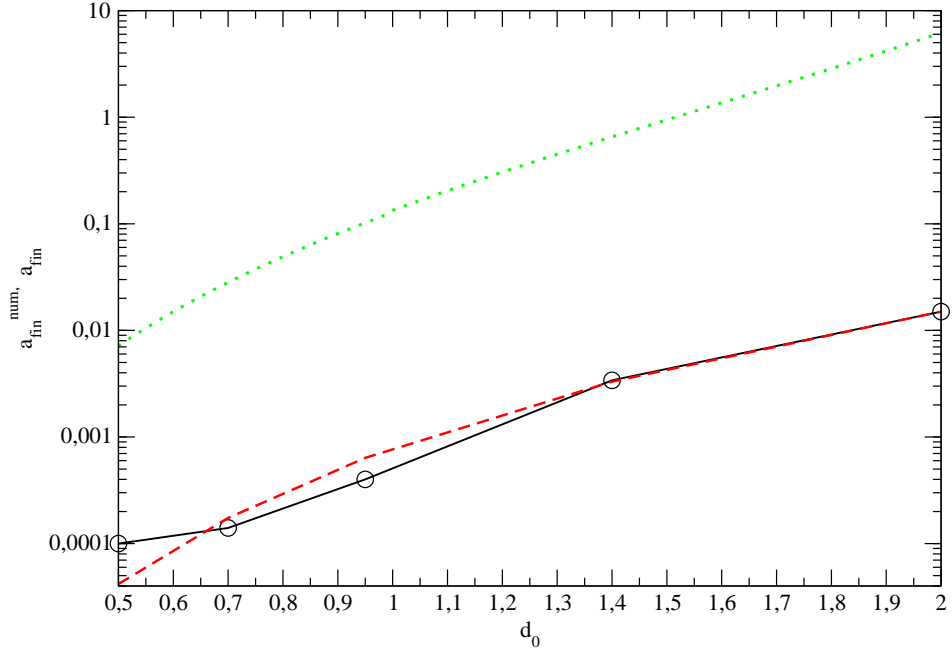}
\caption{We show the values of $\tilde a^{num}_{fin}$ as the solid line, a power law fit of the numerical results by the dashed 
line and the result of the solution of (\ref{e6}) as the dotted line. Symbols corresponds to the values obtained in the different 
numerical runs with different values of $d_{0}$.}
\label{fig5}
\end{figure*}

\begin{figure*}
\includegraphics[width=0.7\textwidth]{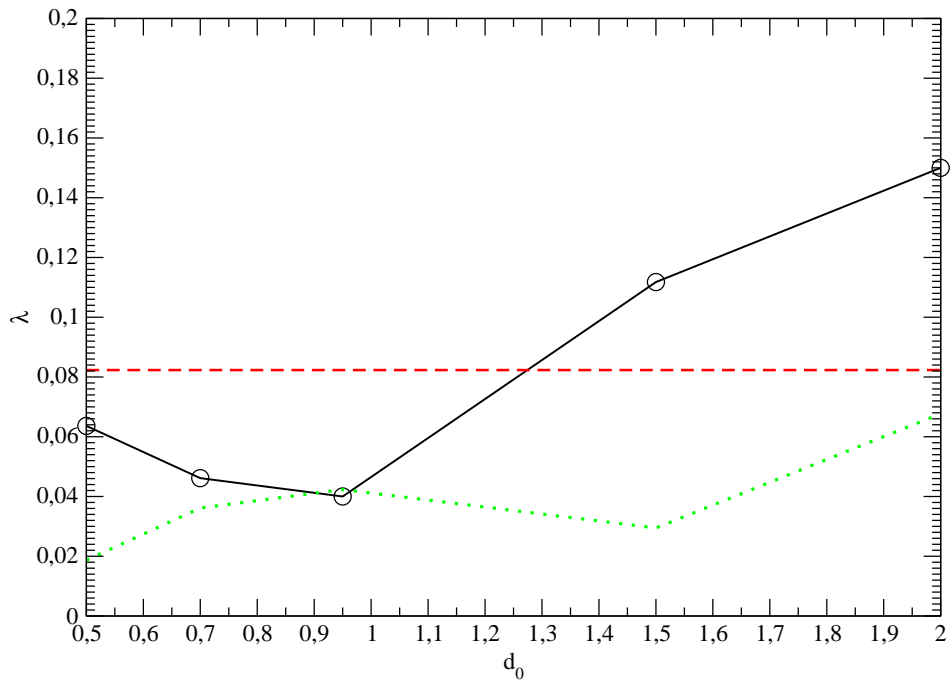}
\caption{The dependency of the quantity $\lambda $ on $d_{0}$ is shown. As in the previous Fig. solid line and symbols correspond 
to the numerical results. The dashed line show the average value of $\lambda$, $\lambda_{av}$ and the dotted line shows the 
deviation $|\lambda-\lambda_{av}|$.}
\label{fig6}
\end{figure*}

We show $\tilde a^{num}_{fin}$ and $\lambda$ in Figs. \ref{fig5} and \ref{fig6}. 
\bigskip
As seen from Fig. \ref{fig5} the numerically obtained values of $\tilde a_{fin}$ are roughly two orders of magnitude smaller than
the analytical ones. However, the slopes of the numerical and analytical curves are similar. The dashed curve shows the power
law fit
\begin{equation}
{\tilde a}^{num}_{fin}=7.9\cdot 10^{-4}d^{4.24},  
\label{e14}
\end{equation}
which gives an excellent agreement with the data apart from the case with the smallest $d_{0}=0.5$, where it 
disagrees by factor of two.

\bigskip
         
In Fig. \ref{fig6} we show the dependency of $\lambda$. 
One can see that $\lambda$ depends non-monotonically on $d_{0}$. However, its average value, $\lambda_{av}=8.23\cdot 10^{-2}$ 
approximates the data points with the accuracy order of or less than $50$ per cent for all runs.

\subsubsection{An estimate of the merger rate}

Given the complexity of the problem on hand we can consider an evaluation of the merger event rate based on the numerical results as a preliminary estimate. Since from Fig. \ref{fig1} it follows that we can use the analytic expression for $a_{fin}$ given by equation (\ref{e8}) for a 
qualitative estimates when $d_0 < 0$, we are going to use (\ref{e14}) and $\lambda_{av}$ and
 all other ingredients needed for the estimate are simple functions of the parameters of the problem,  we can make
it analytically. For that we proceed as follows.

At first, from equations (\ref{e8}), (\ref{e10}) and (\ref{e11}) it is seen that the dimensionless value of angular momentum after the accretion
stage can be estimated as 
\begin{equation}
l_{fin}=\lambda_{av}j_{bh}\sqrt{\tilde a_{fin}}\approx 1.32\cdot 10^{-2}fd^{5}\Lambda,  
\label{e15}
\end{equation}
in case when $f > f_{crit}$ and
\begin{equation}
l_{fin}=\lambda_{av}j_{cp}\sqrt{\tilde a_{fin}}\approx 6.47\cdot 10^{-5}d^{5},  
\label{e16}
\end{equation}
in the opposite case, while from the data represented in Fig. (\ref{fig3}) it follows that
$f_{crit}\approx 6\cdot 10^{-4}$. Next, we express (\ref{e12}) in terms of $l_{fin}$ and ${\tilde a}^{num}_{fin}$ as
\begin{equation}
\tilde t_{GW}=5.22\cdot 10^{19}l_{fin}^{7}\sqrt {(\tilde a^{num}_{fin})}M_{*}^{-5/3}.  
\label{e17}
\end{equation} 
Now we can substitute (\ref{e14}), (\ref{e15}) and (\ref{e16}) into (\ref{e17}) and, using the condition $\tilde t_{GW}=1$ obtain a
value of $d_0$, $d_*$, corresponding to the pairs, which are merging at the present time.     
  
We have 
\begin{equation}
d_{*}=0.735\cdot {(\Lambda f)}^{-0.188}M_{*}^{0.048}  
\label{e18}
\end{equation} 
when $f > f_{crit}$ and 
\begin{equation}
d_{*}=2M_{*}^{0.048}.  
\label{e19}
\end{equation} 
Note that equation (\ref{e18}) give, strictly speaking, an implicit expression for $d_*$, since $\Lambda$ depends on $d_{*}$ itself, 
see equation (\ref{e10}). However, it is also seen that the dependency of $\Lambda$ in (\ref{e18}) is quite weak, and $\Lambda $ 
depends on $d_*$ logarithmically. Therefore, for our estimate below we are going to set $d=1$ in (\ref{e10}) and use 
$\Lambda = \ln \frac{4}{f}$. Also note that from equations (\ref{e18}) and (\ref{e19}) it follows that $d_{*}$ is expected to 
be order of unity.

Since the dependency of $t_{GW}$ on $d$ is close to the case of no accretion, we are going again to use
$\frac{t_{GW}}{X_{*}}\frac{\partial X_{*}}{\partial t_{GW}}=\frac{3}{37}$
in (\ref{en2}), thus obtaining
\begin{equation}
R=2.13\cdot 10^6f^{2}d_{*}^3M_{*}^{-1}{Gpc}^{-3}{yr}^{-1}.  
\label{e20}
\end{equation} 
We substitute (\ref{e18}) and (\ref{e19}) in (\ref{e20}) and obtain
\begin{equation}
R=0.91\cdot 10^6f^{1.44}\Lambda^{-0.564}M_{*}^{-0.856}{Gpc}^{-3}{yr}^{-1}  
\label{e21}
\end{equation} 
and
\begin{equation}
R=1.82\cdot 10^7f^{2}M_{*}^{-0.856}{Gpc}^{-3}{yr}^{-1},  
\label{e22}
\end{equation} 
for the cases $f > f_{crit}$ and $f < f_{crit}$, respectively.

\begin{figure*}
\includegraphics[width=0.7\textwidth]{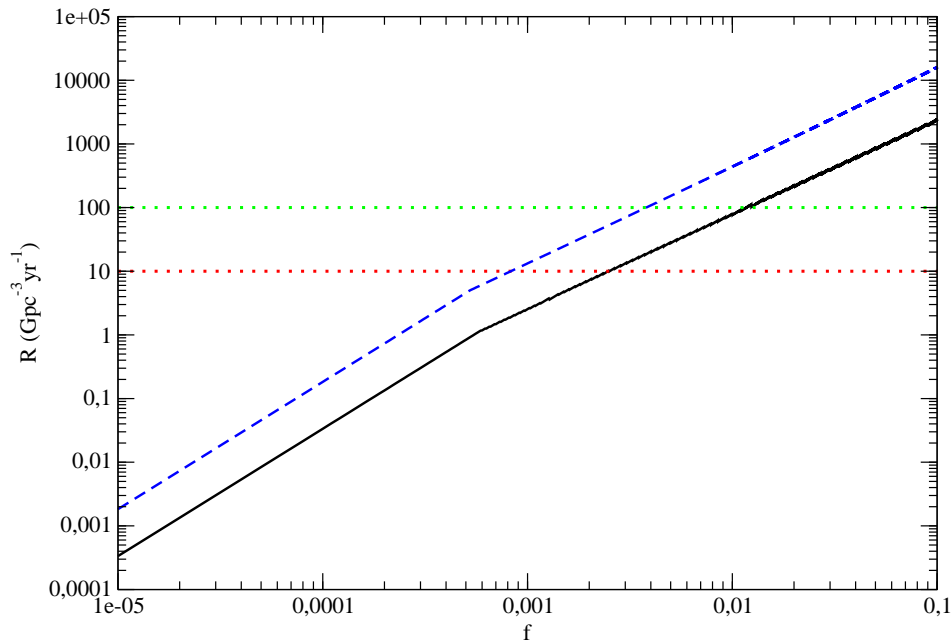}
\caption{The merger rate as a function of PBH mass fraction $f$ for $M_*=1$. The solid curve shows the rate 
obtained under the assumption that CDM accretion effects are negligible, the dashed line shows 
our estimate given by eqs. (\ref{e21}) and (\ref{e22}), which takes into account the effects of 
accretion. The horizontal dotted lines 
show the values $R=10$ and $100$ assumed to be the margins provided by observations.}
\label{fig4}
\end{figure*}          
\bigskip

We show results of evaluation of the event rate in Fig. \ref{fig4}. The case when accretion is neglected is shown as solid line, 
while the case when the effects of accretion are accounted for is represented by dashed line. From this Fig. it is seen that both cases
have approximately the same dependency on $f$, but the latter estimate gives values of $R$ larger than the ones given by the former,
in approximately $6-8$ times. Therefore, in order to satisfy the observational limits on the merger rate, $f$ should be smaller
in the case when the effects of accretion are taken into account. Namely, in this case it should be in the range $7.5\cdot 10^{-4}-3.4\cdot 10^{-3}$ contrary 
to the the standard case when the corresponding range is $2.5\cdot 10^{-3}-1.2\cdot 10^{-2}$.

\section{Conclusions and Discussion}
\label{disc}

In this Paper we consider the influence of interaction  of  CDM particles  with pairs of PBHs immersed in a medium consisting of radiation and CDM  experiencing
the standard cosmological expansion. The pairs become bound at
times comparable to the equipartition time, $t_{eq}$, when both CDM and radiation matter densities are equal to $\rho_{eq}$.  We run simulations with a direct summation N-body code and find numerically that a halo of CDM particles having small angular 
momenta is formed around the pair after it has became bound, with mass $M_{halo}$ growing approximately proportional to the scale factor, $M_{halo} \propto s$,  and the density distribution following the power law $\rho \propto r^{-9/4}$, confirming earlier analytical predictions.  The density distribution is almost stationary at times $t \geq t_{eq}$. Before the formation of the bound pair, provided that the initial comoving separation is large enough, two separate massive halos are formed around two 
black holes. These halos are then destructed during the formation time of the bound pair. We call the bound pairs as 'binary PBHs' and consider values of their parameters (the mass $M=50M_{\odot}$ and initial comoving separations, $d_0$, expressed in unit of the radius $\bar R ={(\frac{3M}{4\pi\rho_{eq}})}^{1/3}$), which are of interest for the problem of the formation of the sources of gravitational waves due to merger of PBH pairs at the present time, possibly explaining the LIGO events. The initial mass fraction 
of PBHs, $f$, is assumed to be small. 

Since the angular momentum of CDM particles residing in the halos is low, they can come close to the binary and gravitationally
interact with it in an efficient way, causing an exchange of energy and angular momentum between the particles and the binary.
It is found numerically, that, as a result of these interactions the binary PBH significantly reduce values of its semimajor axis and angular momentum of the binary. We calculate the evolution of semimajor and angular momentum, taking 
also into account tidal field caused by the presence of radiation matter density. The radiation is assumed to be unclustered and
its matter density obeys the standard cosmological evolution law. Our calculations are started at the epoch of radiation dominated and ended sufficiently deep into the matter-dominated stage. 

We consider several values of $d_0$ in the range $0.5-2$ and a value of the scale factor corresponding to the end time of our calculations, typically ten times larger than that corresponds to $t_{eq}$. We find that, at the end time of our calculations,
the binary angular momentum is typically ten times smaller than its initial value, while a value of semimajor axis is typically
one hundred times smaller than the one obtained in the calculations, which do not take into account the effect of CDM clustering.
Physical origin of these changes seems, however, appears to be different for the binaries having $d_{0} >1$ and $d_{0} < 1$. In the former case it is mainly caused by strong non-stationary processes occurring during the formation time of the bound pair. During this
time a large value of mass of CDM particles is expelled from the system, causing rather abrupt changes of semimajor axis and
angular momentum. In the latter case the evolution mainly proceeds in a secular way, and may be explained by the process of hardening
of the binary due to the interaction with CDM particles. We provide a semi-analytical treatment of this process, which is in an excellent agreement with the numerical results at late times. 

The reason why purely analytical methods appear to be difficult to implement for the problem on hand is because of  relatively rather
large values of angular momentum of the halo's particles. In our setting of the problem the only physical origin of the angular
momentum is provided by torques acting on the particles from the side of quadruple component of the gravitational field of 
the pair. We have checked, however, that these torques alone cannot produce values of angular momentum observed in the simulations.
On the other hand, we have checked than these values cannot be explained by numerical effects, using simulations with a different 
total number of low mass particles, representing CDM, and comparing results of the simulations of a pair with an auxiliary simulation having  a single PBH of doubled mass. We propose the following qualitative picture for an explanation of this issue.
The torques provided by the pair excite non-axisymmetric perturbations of matter density of CDM particles. These perturbations
are amplified in the halo as a results of various potential instabilities, e.g., the instability of radial orbits, see e.g. \cite{2017MNRAS.470.2190P}. Gravitational
field associated with the amplified perturbations causes the torque large enough to explain the obtained values of angular momentum.
Clearly, this picture requires a further detailed analysis.

We use the values of angular momentum and semimajor axis obtained at the end time of our simulations to demonstrate to what
extent the effects considered in this Paper could influence the merger rate. For that, at first, we simplify the approach developed 
in \cite{AKK} in a way, which can be directly used in both case when the CDM clustering is neglected (the standard case) and when it is taken into account. We show, that, regardless of these simplifications, our way to estimate the merger rate quantitatively reproduces the results reported in \cite{AKK} in the case of no clustering. In the case when the clustering is considered the merger rate is shown to be $6-8$ times larger than in the standard case. This, in turn, means that values of $f$ needed to satisfy the
observational limits on the merger rate are several times smaller than in the standard case. We also provide a simple analytical
expressions for the merger rate, which can be directly used in case of a distributed mass spectrum of PBHs.

Our results should be considered as preliminary ones. For example, the process of hardening does not stop at the end time of our 
simulation. It could, potentially, operate up to much later times, thus further decreasing semimajor axis and angular momentum. This, would, obviously, mean even larger values of the merger rate at a fixed value of $f$. A physically motivated end time of 
this process may, in principle, be the time of a significantly non-linear structure formation at the scales of interest, corresponding to redshifts order of ten. On the other hand, halo may become more perturbed at late times due to a slow growth 
of various instabilities and input of the factors neglected in this work (e.g., the influence of other PBHs, development of the halo's perturbations seeded by the cosmological ones). These processes may quench binary's hardening at late 
times. Another interesting development is associated with the process of accretion of baryonic matter, which may
be significant at redshifts order of ten, see e.g. \cite{DeLuca_20, Hasinger_20}. This process may also contribute to an additional significant change of
the orbital parameters of the binary, see \cite{deluca20} for analytic estimates of the significance of
this process. All these issues as well as the issue of relatively large angular momenta of the halo's 
particles are left for a potential future work. 

Finally, we note that simulating PBH pairs with CDM to much larger times is a numerical challenge. Direct summation methods are too slow, while approximate methods, like a tree algorithm used in \texttt{Gadget-2} or \texttt{Gadget-4} code do not provide good enough accuracy, according to our tests.

\begin{acknowledgments}
This work was supported in part by RFBR grant 19-02-00199. We are grateful to
Maxim Pshirkov for very useful comments.

\end{acknowledgments}
\def\jcap{\ref@jnl{J. Cosmology Astropart. Phys.}}
\def\mnras{\ref@jnl{MNRAS}} 
\bibliography{pbh_dm}

\providecommand{\noopsort}[1]{}\providecommand{\singleletter}[1]{#1}%
\begin{thebibliography}{54}%
\makeatletter
\providecommand \@ifxundefined [1]{%
 \@ifx{#1\undefined}
}%
\providecommand \@ifnum [1]{%
 \ifnum #1\expandafter \@firstoftwo
 \else \expandafter \@secondoftwo
 \fi
}%
\providecommand \@ifx [1]{%
 \ifx #1\expandafter \@firstoftwo
 \else \expandafter \@secondoftwo
 \fi
}%
\providecommand \natexlab [1]{#1}%
\providecommand \enquote  [1]{``#1''}%
\providecommand \bibnamefont  [1]{#1}%
\providecommand \bibfnamefont [1]{#1}%
\providecommand \citenamefont [1]{#1}%
\providecommand \href@noop [0]{\@secondoftwo}%
\providecommand \href [0]{\begingroup \@sanitize@url \@href}%
\providecommand \@href[1]{\@@startlink{#1}\@@href}%
\providecommand \@@href[1]{\endgroup#1\@@endlink}%
\providecommand \@sanitize@url [0]{\catcode `\\12\catcode `\$12\catcode
  `\&12\catcode `\#12\catcode `\^12\catcode `\_12\catcode `\%12\relax}%
\providecommand \@@startlink[1]{}%
\providecommand \@@endlink[0]{}%
\providecommand \url  [0]{\begingroup\@sanitize@url \@url }%
\providecommand \@url [1]{\endgroup\@href {#1}{\urlprefix }}%
\providecommand \urlprefix  [0]{URL }%
\providecommand \Eprint [0]{\href }%
\providecommand \doibase [0]{https://doi.org/}%
\providecommand \selectlanguage [0]{\@gobble}%
\providecommand \bibinfo  [0]{\@secondoftwo}%
\providecommand \bibfield  [0]{\@secondoftwo}%
\providecommand \translation [1]{[#1]}%
\providecommand \BibitemOpen [0]{}%
\providecommand \bibitemStop [0]{}%
\providecommand \bibitemNoStop [0]{.\EOS\space}%
\providecommand \EOS [0]{\spacefactor3000\relax}%
\providecommand \BibitemShut  [1]{\csname bibitem#1\endcsname}%
\let\auto@bib@innerbib\@empty
\bibitem [{\citenamefont {{Hawking}}(1971)}]{dm0}%
  \BibitemOpen
  \bibfield  {author} {\bibinfo {author} {\bibfnamefont {S.}~\bibnamefont
  {{Hawking}}},\ }\bibfield  {title} {\bibinfo {title} {{Gravitationally
  collapsed objects of very low mass}},\ }\href
  {https://doi.org/10.1093/mnras/152.1.75} {\bibfield  {journal} {\bibinfo
  {journal} {Monthly Notices of the Royal Astronomical Society}\ }\textbf
  {\bibinfo {volume} {152}},\ \bibinfo {pages} {75} (\bibinfo {year}
  {1971})}\BibitemShut {NoStop}%
\bibitem [{\citenamefont {Lacki}\ and\ \citenamefont {Beacom}(2010)}]{dm1}%
  \BibitemOpen
  \bibfield  {author} {\bibinfo {author} {\bibfnamefont {B.~C.}\ \bibnamefont
  {Lacki}}\ and\ \bibinfo {author} {\bibfnamefont {J.~F.}\ \bibnamefont
  {Beacom}},\ }\bibfield  {title} {\bibinfo {title} {{PRIMORDIAL} {BLACK}
  {HOLES} {AS} {DARK} {MATTER}: {ALMOST} {ALL} {OR} {ALMOST} {NOTHING}},\
  }\href {https://doi.org/10.1088/2041-8205/720/1/l67} {\bibfield  {journal}
  {\bibinfo  {journal} {The Astrophysical Journal}\ }\textbf {\bibinfo {volume}
  {720}},\ \bibinfo {pages} {L67} (\bibinfo {year} {2010})}\BibitemShut
  {NoStop}%
\bibitem [{\citenamefont {Belotsky}\ \emph {et~al.}(2014)\citenamefont
  {Belotsky}, \citenamefont {Dmitriev}, \citenamefont {Esipova}, \citenamefont
  {Gani}, \citenamefont {Grobov}, \citenamefont {Khlopov}, \citenamefont
  {Kirillov}, \citenamefont {Rubin},\ and\ \citenamefont {Svadkovsky}}]{dm2}%
  \BibitemOpen
  \bibfield  {author} {\bibinfo {author} {\bibfnamefont {K.~M.}\ \bibnamefont
  {Belotsky}}, \bibinfo {author} {\bibfnamefont {A.~E.}\ \bibnamefont
  {Dmitriev}}, \bibinfo {author} {\bibfnamefont {E.~A.}\ \bibnamefont
  {Esipova}}, \bibinfo {author} {\bibfnamefont {V.~A.}\ \bibnamefont {Gani}},
  \bibinfo {author} {\bibfnamefont {A.~V.}\ \bibnamefont {Grobov}}, \bibinfo
  {author} {\bibfnamefont {M.~Y.}\ \bibnamefont {Khlopov}}, \bibinfo {author}
  {\bibfnamefont {A.~A.}\ \bibnamefont {Kirillov}}, \bibinfo {author}
  {\bibfnamefont {S.~G.}\ \bibnamefont {Rubin}},\ and\ \bibinfo {author}
  {\bibfnamefont {I.~V.}\ \bibnamefont {Svadkovsky}},\ }\bibfield  {title}
  {\bibinfo {title} {Signatures of primordial black hole dark matter},\ }\href
  {https://doi.org/10.1142/s0217732314400057} {\bibfield  {journal} {\bibinfo
  {journal} {Modern Physics Letters A}\ }\textbf {\bibinfo {volume} {29}},\
  \bibinfo {pages} {1440005} (\bibinfo {year} {2014})}\BibitemShut {NoStop}%
\bibitem [{\citenamefont {Kashlinsky}(2016)}]{dm3}%
  \BibitemOpen
  \bibfield  {author} {\bibinfo {author} {\bibfnamefont {A.}~\bibnamefont
  {Kashlinsky}},\ }\bibfield  {title} {\bibinfo {title} {{LIGO} {GRAVITATIONAL}
  {WAVE} {DETECTION}, {PRIMORDIAL} {BLACK} {HOLES}, {AND} {THE} {NEAR}-{IR}
  {COSMIC} {INFRARED} {BACKGROUND} {ANISOTROPIES}},\ }\href
  {https://doi.org/10.3847/2041-8205/823/2/l25} {\bibfield  {journal} {\bibinfo
   {journal} {The Astrophysical Journal}\ }\textbf {\bibinfo {volume} {823}},\
  \bibinfo {pages} {L25} (\bibinfo {year} {2016})}\BibitemShut {NoStop}%
\bibitem [{\citenamefont {Clesse}\ and\ \citenamefont
  {García-Bellido}(2017)}]{dm4}%
  \BibitemOpen
  \bibfield  {author} {\bibinfo {author} {\bibfnamefont {S.}~\bibnamefont
  {Clesse}}\ and\ \bibinfo {author} {\bibfnamefont {J.}~\bibnamefont
  {García-Bellido}},\ }\bibfield  {title} {\bibinfo {title} {The clustering of
  massive primordial black holes as dark matter: Measuring their mass
  distribution with advanced ligo},\ }\href
  {https://doi.org/10.1016/j.dark.2016.10.002} {\bibfield  {journal} {\bibinfo
  {journal} {Physics of the Dark Universe}\ }\textbf {\bibinfo {volume} {15}},\
  \bibinfo {pages} {142–147} (\bibinfo {year} {2017})}\BibitemShut {NoStop}%
\bibitem [{\citenamefont {{Clesse}}\ and\ \citenamefont
  {{García-Bellido}}(2018)}]{dm5}%
  \BibitemOpen
  \bibfield  {author} {\bibinfo {author} {\bibfnamefont {S.}~\bibnamefont
  {{Clesse}}}\ and\ \bibinfo {author} {\bibfnamefont {J.}~\bibnamefont
  {{García-Bellido}}},\ }\bibfield  {title} {\bibinfo {title} {{Seven hints
  for primordial black hole dark matter}},\ }\href
  {https://doi.org/10.1016/j.dark.2018.08.004} {\bibfield  {journal} {\bibinfo
  {journal} {Physics of the Dark Universe}\ }\textbf {\bibinfo {volume} {22}},\
  \bibinfo {pages} {137} (\bibinfo {year} {2018})},\ \Eprint
  {https://arxiv.org/abs/1711.10458} {arXiv:1711.10458 [astro-ph.CO]}
  \BibitemShut {NoStop}%
\bibitem [{\citenamefont {Espinosa}\ \emph {et~al.}(2018)\citenamefont
  {Espinosa}, \citenamefont {Racco},\ and\ \citenamefont {Riotto}}]{dm6}%
  \BibitemOpen
  \bibfield  {author} {\bibinfo {author} {\bibfnamefont {J.~R.}\ \bibnamefont
  {Espinosa}}, \bibinfo {author} {\bibfnamefont {D.}~\bibnamefont {Racco}},\
  and\ \bibinfo {author} {\bibfnamefont {A.}~\bibnamefont {Riotto}},\
  }\bibfield  {title} {\bibinfo {title} {Cosmological signature of the standard
  model higgs vacuum instability: Primordial black holes as dark matter},\
  }\href {https://doi.org/10.1103/PhysRevLett.120.121301} {\bibfield  {journal}
  {\bibinfo  {journal} {Phys. Rev. Lett.}\ }\textbf {\bibinfo {volume} {120}},\
  \bibinfo {pages} {121301} (\bibinfo {year} {2018})}\BibitemShut {NoStop}%
\bibitem [{\citenamefont {Carr}\ and\ \citenamefont {Rees}(1984)}]{smbh_seed1}%
  \BibitemOpen
  \bibfield  {author} {\bibinfo {author} {\bibfnamefont {B.~J.}\ \bibnamefont
  {Carr}}\ and\ \bibinfo {author} {\bibfnamefont {M.~J.}\ \bibnamefont
  {Rees}},\ }\bibfield  {title} {\bibinfo {title} {{Can pregalactic objects
  generate galaxies?}},\ }\href {https://doi.org/10.1093/mnras/206.4.801}
  {\bibfield  {journal} {\bibinfo  {journal} {Monthly Notices of the Royal
  Astronomical Society}\ }\textbf {\bibinfo {volume} {206}},\ \bibinfo {pages}
  {801} (\bibinfo {year} {1984})},\ \Eprint
  {https://arxiv.org/abs/https://academic.oup.com/mnras/article-pdf/206/4/801/2902772/mnras206-0801.pdf}
  {https://academic.oup.com/mnras/article-pdf/206/4/801/2902772/mnras206-0801.pdf}
  \BibitemShut {NoStop}%
\bibitem [{\citenamefont {Bean}\ and\ \citenamefont
  {Magueijo}(2002)}]{smbh_seed2}%
  \BibitemOpen
  \bibfield  {author} {\bibinfo {author} {\bibfnamefont {R.}~\bibnamefont
  {Bean}}\ and\ \bibinfo {author} {\bibfnamefont {J.~a.}\ \bibnamefont
  {Magueijo}},\ }\bibfield  {title} {\bibinfo {title} {Could supermassive black
  holes be quintessential primordial black holes?},\ }\href
  {https://doi.org/10.1103/PhysRevD.66.063505} {\bibfield  {journal} {\bibinfo
  {journal} {Phys. Rev. D}\ }\textbf {\bibinfo {volume} {66}},\ \bibinfo
  {pages} {063505} (\bibinfo {year} {2002})}\BibitemShut {NoStop}%
\bibitem [{\citenamefont {{Polnarev}}\ and\ \citenamefont
  {{Khlopov}}(1985)}]{polnarev85}%
  \BibitemOpen
  \bibfield  {author} {\bibinfo {author} {\bibfnamefont {A.~G.}\ \bibnamefont
  {{Polnarev}}}\ and\ \bibinfo {author} {\bibfnamefont {M.~Y.}\ \bibnamefont
  {{Khlopov}}},\ }\bibfield  {title} {\bibinfo {title} {{REVIEWS OF TOPICAL
  PROBLEMS: Cosmology, primordial black holes, and supermassive particles}},\
  }\href {https://doi.org/10.1070/PU1985v028n03ABEH003858} {\bibfield
  {journal} {\bibinfo  {journal} {Soviet Physics Uspekhi}\ }\textbf {\bibinfo
  {volume} {28}},\ \bibinfo {pages} {213} (\bibinfo {year} {1985})}\BibitemShut
  {NoStop}%
\bibitem [{\citenamefont {{Carr}}\ and\ \citenamefont
  {{K{\"u}hnel}}(2020)}]{Carr_20_candi}%
  \BibitemOpen
  \bibfield  {author} {\bibinfo {author} {\bibfnamefont {B.}~\bibnamefont
  {{Carr}}}\ and\ \bibinfo {author} {\bibfnamefont {F.}~\bibnamefont
  {{K{\"u}hnel}}},\ }\bibfield  {title} {\bibinfo {title} {{Primordial Black
  Holes as Dark Matter: Recent Developments}},\ }\href
  {https://doi.org/10.1146/annurev-nucl-050520-125911} {\bibfield  {journal}
  {\bibinfo  {journal} {Annual Review of Nuclear and Particle Science}\
  }\textbf {\bibinfo {volume} {70}},\ \bibinfo {pages} {355} (\bibinfo {year}
  {2020})},\ \Eprint {https://arxiv.org/abs/2006.02838} {arXiv:2006.02838
  [astro-ph.CO]} \BibitemShut {NoStop}%
\bibitem [{\citenamefont {{Carr}}\ and\ \citenamefont
  {{Kuhnel}}(2021)}]{Carr_21_candi}%
  \BibitemOpen
  \bibfield  {author} {\bibinfo {author} {\bibfnamefont {B.}~\bibnamefont
  {{Carr}}}\ and\ \bibinfo {author} {\bibfnamefont {F.}~\bibnamefont
  {{Kuhnel}}},\ }\bibfield  {title} {\bibinfo {title} {{Primordial Black Holes
  as Dark Matter Candidates}},\ }\href@noop {} {\bibfield  {journal} {\bibinfo
  {journal} {arXiv e-prints}\ ,\ \bibinfo {eid} {arXiv:2110.02821}} (\bibinfo
  {year} {2021})},\ \Eprint {https://arxiv.org/abs/2110.02821}
  {arXiv:2110.02821 [astro-ph.CO]} \BibitemShut {NoStop}%
\bibitem [{\citenamefont {{Abbott}}\ \emph
  {et~al.}(2016{\natexlab{a}})\citenamefont {{Abbott}}, \citenamefont
  {{Abbott}}, \citenamefont {{Abbott}}, \citenamefont {{Abernathy}},
  \citenamefont {{Acernese}}, \citenamefont {{Ackley}}, \citenamefont
  {{Adams}}, \citenamefont {{Adams}}, \citenamefont {{Addesso}},\ and\
  \citenamefont {{Adhikari}}}]{ligo1}%
  \BibitemOpen
  \bibfield  {author} {\bibinfo {author} {\bibfnamefont {B.~P.}\ \bibnamefont
  {{Abbott}}}, \bibinfo {author} {\bibfnamefont {R.}~\bibnamefont {{Abbott}}},
  \bibinfo {author} {\bibfnamefont {T.~D.}\ \bibnamefont {{Abbott}}}, \bibinfo
  {author} {\bibfnamefont {M.~R.}\ \bibnamefont {{Abernathy}}}, \bibinfo
  {author} {\bibfnamefont {F.}~\bibnamefont {{Acernese}}}, \bibinfo {author}
  {\bibfnamefont {K.}~\bibnamefont {{Ackley}}}, \bibinfo {author}
  {\bibfnamefont {C.}~\bibnamefont {{Adams}}}, \bibinfo {author} {\bibfnamefont
  {T.}~\bibnamefont {{Adams}}}, \bibinfo {author} {\bibfnamefont
  {P.}~\bibnamefont {{Addesso}}},\ and\ \bibinfo {author} {\bibfnamefont
  {R.~X.}\ \bibnamefont {{Adhikari}}},\ }\bibfield  {title} {\bibinfo {title}
  {{Observation of Gravitational Waves from a Binary Black Hole Merger}},\
  }\href {https://doi.org/10.1103/PhysRevLett.116.061102} {\bibfield  {journal}
  {\bibinfo  {journal} {\prl}\ }\textbf {\bibinfo {volume} {116}},\ \bibinfo
  {eid} {061102} (\bibinfo {year} {2016}{\natexlab{a}})},\ \Eprint
  {https://arxiv.org/abs/1602.03837} {arXiv:1602.03837 [gr-qc]} \BibitemShut
  {NoStop}%
\bibitem [{\citenamefont {{Abbott}}\ \emph
  {et~al.}(2016{\natexlab{b}})\citenamefont {{Abbott}}, \citenamefont
  {{Abbott}}, \citenamefont {{Abbott}}, \citenamefont {{Abernathy}},
  \citenamefont {{Acernese}}, \citenamefont {{Ackley}}, \citenamefont
  {{Adams}}, \citenamefont {{Adams}}, \citenamefont {{Addesso}},\ and\
  \citenamefont {{Adhikari}}}]{ligo2}%
  \BibitemOpen
  \bibfield  {author} {\bibinfo {author} {\bibfnamefont {B.~P.}\ \bibnamefont
  {{Abbott}}}, \bibinfo {author} {\bibfnamefont {R.}~\bibnamefont {{Abbott}}},
  \bibinfo {author} {\bibfnamefont {T.~D.}\ \bibnamefont {{Abbott}}}, \bibinfo
  {author} {\bibfnamefont {M.~R.}\ \bibnamefont {{Abernathy}}}, \bibinfo
  {author} {\bibfnamefont {F.}~\bibnamefont {{Acernese}}}, \bibinfo {author}
  {\bibfnamefont {K.}~\bibnamefont {{Ackley}}}, \bibinfo {author}
  {\bibfnamefont {C.}~\bibnamefont {{Adams}}}, \bibinfo {author} {\bibfnamefont
  {T.}~\bibnamefont {{Adams}}}, \bibinfo {author} {\bibfnamefont
  {P.}~\bibnamefont {{Addesso}}},\ and\ \bibinfo {author} {\bibfnamefont
  {R.~X.}\ \bibnamefont {{Adhikari}}},\ }\bibfield  {title} {\bibinfo {title}
  {{GW151226: Observation of Gravitational Waves from a 22-Solar-Mass Binary
  Black Hole Coalescence}},\ }\href
  {https://doi.org/10.1103/PhysRevLett.116.241103} {\bibfield  {journal}
  {\bibinfo  {journal} {\prl}\ }\textbf {\bibinfo {volume} {116}},\ \bibinfo
  {eid} {241103} (\bibinfo {year} {2016}{\natexlab{b}})},\ \Eprint
  {https://arxiv.org/abs/1606.04855} {arXiv:1606.04855 [gr-qc]} \BibitemShut
  {NoStop}%
\bibitem [{\citenamefont {{Abbott}}\ \emph
  {et~al.}(2016{\natexlab{c}})\citenamefont {{Abbott}}, \citenamefont
  {{Abbott}}, \citenamefont {{Abbott}}, \citenamefont {{Abernathy}},
  \citenamefont {{Acernese}}, \citenamefont {{Ackley}}, \citenamefont
  {{Adams}}, \citenamefont {{Adams}}, \citenamefont {{Addesso}},\ and\
  \citenamefont {{Adhikari}}}]{ligo_merge}%
  \BibitemOpen
  \bibfield  {author} {\bibinfo {author} {\bibfnamefont {B.~P.}\ \bibnamefont
  {{Abbott}}}, \bibinfo {author} {\bibfnamefont {R.}~\bibnamefont {{Abbott}}},
  \bibinfo {author} {\bibfnamefont {T.~D.}\ \bibnamefont {{Abbott}}}, \bibinfo
  {author} {\bibfnamefont {M.~R.}\ \bibnamefont {{Abernathy}}}, \bibinfo
  {author} {\bibfnamefont {F.}~\bibnamefont {{Acernese}}}, \bibinfo {author}
  {\bibfnamefont {K.}~\bibnamefont {{Ackley}}}, \bibinfo {author}
  {\bibfnamefont {C.}~\bibnamefont {{Adams}}}, \bibinfo {author} {\bibfnamefont
  {T.}~\bibnamefont {{Adams}}}, \bibinfo {author} {\bibfnamefont
  {P.}~\bibnamefont {{Addesso}}},\ and\ \bibinfo {author} {\bibfnamefont
  {R.~X.}\ \bibnamefont {{Adhikari}}},\ }\bibfield  {title} {\bibinfo {title}
  {{Astrophysical Implications of the Binary Black-hole Merger GW150914}},\
  }\href {https://doi.org/10.3847/2041-8205/818/2/L22} {\bibfield  {journal}
  {\bibinfo  {journal} {The Astrophysical Journal}\ }\textbf {\bibinfo {volume}
  {818}},\ \bibinfo {eid} {L22} (\bibinfo {year} {2016}{\natexlab{c}})},\
  \Eprint {https://arxiv.org/abs/1602.03846} {arXiv:1602.03846 [astro-ph.HE]}
  \BibitemShut {NoStop}%
\bibitem [{\citenamefont {{Abbott}}\ \emph
  {et~al.}(2017{\natexlab{a}})\citenamefont {{Abbott}}, \citenamefont
  {{Abbott}}, \citenamefont {{Abbott}}, \citenamefont {{Acernese}},
  \citenamefont {{Ackley}}, \citenamefont {{Adams}}, \citenamefont {{Adams}},
  \citenamefont {{Addesso}}, \citenamefont {{Adhikari}},\ and\ \citenamefont
  {{Adya}}}]{ligo3}%
  \BibitemOpen
  \bibfield  {author} {\bibinfo {author} {\bibfnamefont {B.~P.}\ \bibnamefont
  {{Abbott}}}, \bibinfo {author} {\bibfnamefont {R.}~\bibnamefont {{Abbott}}},
  \bibinfo {author} {\bibfnamefont {T.~D.}\ \bibnamefont {{Abbott}}}, \bibinfo
  {author} {\bibfnamefont {F.}~\bibnamefont {{Acernese}}}, \bibinfo {author}
  {\bibfnamefont {K.}~\bibnamefont {{Ackley}}}, \bibinfo {author}
  {\bibfnamefont {C.}~\bibnamefont {{Adams}}}, \bibinfo {author} {\bibfnamefont
  {T.}~\bibnamefont {{Adams}}}, \bibinfo {author} {\bibfnamefont
  {P.}~\bibnamefont {{Addesso}}}, \bibinfo {author} {\bibfnamefont {R.~X.}\
  \bibnamefont {{Adhikari}}},\ and\ \bibinfo {author} {\bibfnamefont {V.~B.}\
  \bibnamefont {{Adya}}},\ }\bibfield  {title} {\bibinfo {title} {{GW170104:
  Observation of a 50-Solar-Mass Binary Black Hole Coalescence at Redshift
  0.2}},\ }\href {https://doi.org/10.1103/PhysRevLett.118.221101} {\bibfield
  {journal} {\bibinfo  {journal} {\prl}\ }\textbf {\bibinfo {volume} {118}},\
  \bibinfo {eid} {221101} (\bibinfo {year} {2017}{\natexlab{a}})},\ \Eprint
  {https://arxiv.org/abs/1706.01812} {arXiv:1706.01812 [gr-qc]} \BibitemShut
  {NoStop}%
\bibitem [{\citenamefont {{Abbott}}\ \emph
  {et~al.}(2017{\natexlab{b}})\citenamefont {{Abbott}}, \citenamefont
  {{Abbott}}, \citenamefont {{Abbott}}, \citenamefont {{Acernese}},
  \citenamefont {{Ackley}}, \citenamefont {{Adams}}, \citenamefont {{Adams}},
  \citenamefont {{Addesso}}, \citenamefont {{Adhikari}},\ and\ \citenamefont
  {{Adya}}}]{ligo5}%
  \BibitemOpen
  \bibfield  {author} {\bibinfo {author} {\bibfnamefont {B.~P.}\ \bibnamefont
  {{Abbott}}}, \bibinfo {author} {\bibfnamefont {R.}~\bibnamefont {{Abbott}}},
  \bibinfo {author} {\bibfnamefont {T.~D.}\ \bibnamefont {{Abbott}}}, \bibinfo
  {author} {\bibfnamefont {F.}~\bibnamefont {{Acernese}}}, \bibinfo {author}
  {\bibfnamefont {K.}~\bibnamefont {{Ackley}}}, \bibinfo {author}
  {\bibfnamefont {C.}~\bibnamefont {{Adams}}}, \bibinfo {author} {\bibfnamefont
  {T.}~\bibnamefont {{Adams}}}, \bibinfo {author} {\bibfnamefont
  {P.}~\bibnamefont {{Addesso}}}, \bibinfo {author} {\bibfnamefont {R.~X.}\
  \bibnamefont {{Adhikari}}},\ and\ \bibinfo {author} {\bibfnamefont {V.~B.}\
  \bibnamefont {{Adya}}},\ }\bibfield  {title} {\bibinfo {title} {{GW170814: A
  Three-Detector Observation of Gravitational Waves from a Binary Black Hole
  Coalescence}},\ }\href {https://doi.org/10.1103/PhysRevLett.119.141101}
  {\bibfield  {journal} {\bibinfo  {journal} {\prl}\ }\textbf {\bibinfo
  {volume} {119}},\ \bibinfo {eid} {141101} (\bibinfo {year}
  {2017}{\natexlab{b}})},\ \Eprint {https://arxiv.org/abs/1709.09660}
  {arXiv:1709.09660 [gr-qc]} \BibitemShut {NoStop}%
\bibitem [{\citenamefont {{Abbott}}\ \emph
  {et~al.}(2017{\natexlab{c}})\citenamefont {{Abbott}}, \citenamefont
  {{Abbott}}, \citenamefont {{Abbott}}, \citenamefont {{Acernese}},
  \citenamefont {{Ackley}}, \citenamefont {{Adams}}, \citenamefont {{Adams}},
  \citenamefont {{Addesso}}, \citenamefont {{Adhikari}},\ and\ \citenamefont
  {{Adya}}}]{ligo4}%
  \BibitemOpen
  \bibfield  {author} {\bibinfo {author} {\bibfnamefont {B.~P.}\ \bibnamefont
  {{Abbott}}}, \bibinfo {author} {\bibfnamefont {R.}~\bibnamefont {{Abbott}}},
  \bibinfo {author} {\bibfnamefont {T.~D.}\ \bibnamefont {{Abbott}}}, \bibinfo
  {author} {\bibfnamefont {F.}~\bibnamefont {{Acernese}}}, \bibinfo {author}
  {\bibfnamefont {K.}~\bibnamefont {{Ackley}}}, \bibinfo {author}
  {\bibfnamefont {C.}~\bibnamefont {{Adams}}}, \bibinfo {author} {\bibfnamefont
  {T.}~\bibnamefont {{Adams}}}, \bibinfo {author} {\bibfnamefont
  {P.}~\bibnamefont {{Addesso}}}, \bibinfo {author} {\bibfnamefont {R.~X.}\
  \bibnamefont {{Adhikari}}},\ and\ \bibinfo {author} {\bibfnamefont {V.~B.}\
  \bibnamefont {{Adya}}},\ }\bibfield  {title} {\bibinfo {title} {{GW170608:
  Observation of a 19 Solar-mass Binary Black Hole Coalescence}},\ }\href
  {https://doi.org/10.3847/2041-8213/aa9f0c} {\bibfield  {journal} {\bibinfo
  {journal} {\apj}\ }\textbf {\bibinfo {volume} {851}},\ \bibinfo {eid} {L35}
  (\bibinfo {year} {2017}{\natexlab{c}})},\ \Eprint
  {https://arxiv.org/abs/1711.05578} {arXiv:1711.05578 [astro-ph.HE]}
  \BibitemShut {NoStop}%
\bibitem [{\citenamefont {{Abbott}}\ \emph {et~al.}(2020)\citenamefont
  {{Abbott}}, \citenamefont {{Abbott}}, \citenamefont {{Abbott}}, \citenamefont
  {{Acernese}}, \citenamefont {{Ackley}}, \citenamefont {{Adams}},
  \citenamefont {{Adams}}, \citenamefont {{Addesso}}, \citenamefont
  {{Adhikari}},\ and\ \citenamefont {{Adya}}}]{ligo150}%
  \BibitemOpen
  \bibfield  {author} {\bibinfo {author} {\bibfnamefont {B.~P.}\ \bibnamefont
  {{Abbott}}}, \bibinfo {author} {\bibfnamefont {R.}~\bibnamefont {{Abbott}}},
  \bibinfo {author} {\bibfnamefont {T.~D.}\ \bibnamefont {{Abbott}}}, \bibinfo
  {author} {\bibfnamefont {F.}~\bibnamefont {{Acernese}}}, \bibinfo {author}
  {\bibfnamefont {K.}~\bibnamefont {{Ackley}}}, \bibinfo {author}
  {\bibfnamefont {C.}~\bibnamefont {{Adams}}}, \bibinfo {author} {\bibfnamefont
  {T.}~\bibnamefont {{Adams}}}, \bibinfo {author} {\bibfnamefont
  {P.}~\bibnamefont {{Addesso}}}, \bibinfo {author} {\bibfnamefont {R.~X.}\
  \bibnamefont {{Adhikari}}},\ and\ \bibinfo {author} {\bibfnamefont {V.~B.}\
  \bibnamefont {{Adya}}},\ }\bibfield  {title} {\bibinfo {title} {Properties
  and astrophysical implications of the 150 m $\odot$ binary black hole merger
  {GW}190521},\ }\href {https://doi.org/10.3847/2041-8213/aba493} {\bibfield
  {journal} {\bibinfo  {journal} {The Astrophysical Journal}\ }\textbf
  {\bibinfo {volume} {900}},\ \bibinfo {pages} {L13} (\bibinfo {year}
  {2020})}\BibitemShut {NoStop}%
\bibitem [{Note1()}]{Note1}%
  \BibitemOpen
  \bibinfo {note} {LIGO/Virgo Public Alerts from the O3/2019 observational run
  can be found at https://gracedb.ligo.org/superevents/public/O3/}\BibitemShut
  {NoStop}%
\bibitem [{\citenamefont {{Kovetz}}(2017)}]{Kovetz_17}%
  \BibitemOpen
  \bibfield  {author} {\bibinfo {author} {\bibfnamefont {E.~D.}\ \bibnamefont
  {{Kovetz}}},\ }\bibfield  {title} {\bibinfo {title} {{Probing Primordial
  Black Hole Dark Matter with Gravitational Waves}},\ }\href
  {https://doi.org/10.1103/PhysRevLett.119.131301} {\bibfield  {journal}
  {\bibinfo  {journal} {Physical Review Letters}\ }\textbf {\bibinfo {volume}
  {119}},\ \bibinfo {eid} {131301} (\bibinfo {year} {2017})},\ \Eprint
  {https://arxiv.org/abs/1705.09182} {arXiv:1705.09182 [astro-ph.CO]}
  \BibitemShut {NoStop}%
\bibitem [{\citenamefont {{Sasaki}}\ \emph {et~al.}(2016)\citenamefont
  {{Sasaki}}, \citenamefont {{Suyama}}, \citenamefont {{Tanaka}},\ and\
  \citenamefont {{Yokoyama}}}]{Sasaki_16}%
  \BibitemOpen
  \bibfield  {author} {\bibinfo {author} {\bibfnamefont {M.}~\bibnamefont
  {{Sasaki}}}, \bibinfo {author} {\bibfnamefont {T.}~\bibnamefont {{Suyama}}},
  \bibinfo {author} {\bibfnamefont {T.}~\bibnamefont {{Tanaka}}},\ and\
  \bibinfo {author} {\bibfnamefont {S.}~\bibnamefont {{Yokoyama}}},\ }\bibfield
   {title} {\bibinfo {title} {{Primordial Black Hole Scenario for the
  Gravitational-Wave Event GW150914}},\ }\href
  {https://doi.org/10.1103/PhysRevLett.117.061101} {\bibfield  {journal}
  {\bibinfo  {journal} {Physical Review Letters}\ }\textbf {\bibinfo {volume}
  {117}},\ \bibinfo {eid} {061101} (\bibinfo {year} {2016})},\ \Eprint
  {https://arxiv.org/abs/1603.08338} {arXiv:1603.08338 [astro-ph.CO]}
  \BibitemShut {NoStop}%
\bibitem [{\citenamefont {{Blinnikov}}\ \emph {et~al.}(2016)\citenamefont
  {{Blinnikov}}, \citenamefont {{Dolgov}}, \citenamefont {{Porayko}},\ and\
  \citenamefont {{Postnov}}}]{Blinnikov_16}%
  \BibitemOpen
  \bibfield  {author} {\bibinfo {author} {\bibfnamefont {S.}~\bibnamefont
  {{Blinnikov}}}, \bibinfo {author} {\bibfnamefont {A.}~\bibnamefont
  {{Dolgov}}}, \bibinfo {author} {\bibfnamefont {N.~K.}\ \bibnamefont
  {{Porayko}}},\ and\ \bibinfo {author} {\bibfnamefont {K.}~\bibnamefont
  {{Postnov}}},\ }\bibfield  {title} {\bibinfo {title} {{Solving puzzles of
  GW150914 by primordial black holes}},\ }\href
  {https://doi.org/10.1088/1475-7516/2016/11/036} {\bibfield  {journal}
  {\bibinfo  {journal} {Journal of Cosmology and Astroparticle Physics}\
  }\textbf {\bibinfo {volume} {2016}}\bibfield  {number} {\bibinfo  {number} {
  (11)},\ \bibinfo {eid} {036}},\ }\Eprint {https://arxiv.org/abs/1611.00541}
  {arXiv:1611.00541 [astro-ph.HE]} \BibitemShut {NoStop}%
\bibitem [{\citenamefont {{Mukherjee}}\ and\ \citenamefont
  {{Silk}}(2021)}]{Mukherjee21}%
  \BibitemOpen
  \bibfield  {author} {\bibinfo {author} {\bibfnamefont {S.}~\bibnamefont
  {{Mukherjee}}}\ and\ \bibinfo {author} {\bibfnamefont {J.}~\bibnamefont
  {{Silk}}},\ }\bibfield  {title} {\bibinfo {title} {{Can we distinguish
  astrophysical from primordial black holes via the stochastic gravitational
  wave background?}},\ }\href {https://doi.org/10.1093/mnras/stab1932}
  {\bibfield  {journal} {\bibinfo  {journal} {\mnras}\ }\textbf {\bibinfo
  {volume} {506}},\ \bibinfo {pages} {3977} (\bibinfo {year} {2021})},\ \Eprint
  {https://arxiv.org/abs/2105.11139} {arXiv:2105.11139 [gr-qc]} \BibitemShut
  {NoStop}%
\bibitem [{Note2()}]{Note2}%
  \BibitemOpen
  \bibinfo {note} {Note, however, that some of the LIGO/Virgo events may have
  another origin, see e.g. \cite {astro_bh}. It is important to stress that
  even if all LIGO/Virgo events will be shown to be due to mergers of black
  holes of stellar origin, if would give a very non-trivial constraint on the
  abundance of PBHs of stellar masses in the Universe provided there is a
  reliable way of an estimate of the formation of PBHs pairs with orbital
  parameters appropriate for the LIGO/Virgo events for a given abundance of
  PBH. The exact possible mass fraction of PBHs in dark matter (DM) in
  LIGO/Virgo mass range is still a subject of large theoretical uncertainties,
  see, e.g. \cite {constr_war1, constr_war2}.}\BibitemShut {Stop}%
\bibitem [{\citenamefont {{Nakamura}}\ \emph {et~al.}(1997)\citenamefont
  {{Nakamura}}, \citenamefont {{Sasaki}}, \citenamefont {{Tanaka}},\ and\
  \citenamefont {{Thorne}}}]{Nakamura_97}%
  \BibitemOpen
  \bibfield  {author} {\bibinfo {author} {\bibfnamefont {T.}~\bibnamefont
  {{Nakamura}}}, \bibinfo {author} {\bibfnamefont {M.}~\bibnamefont
  {{Sasaki}}}, \bibinfo {author} {\bibfnamefont {T.}~\bibnamefont {{Tanaka}}},\
  and\ \bibinfo {author} {\bibfnamefont {K.~S.}\ \bibnamefont {{Thorne}}},\
  }\bibfield  {title} {\bibinfo {title} {{Gravitational Waves from Coalescing
  Black Hole MACHO Binaries}},\ }\href {https://doi.org/10.1086/310886}
  {\bibfield  {journal} {\bibinfo  {journal} {The Astrophysical Journal}\
  }\textbf {\bibinfo {volume} {487}},\ \bibinfo {pages} {L139} (\bibinfo {year}
  {1997})},\ \Eprint {https://arxiv.org/abs/astro-ph/9708060}
  {arXiv:astro-ph/9708060 [astro-ph]} \BibitemShut {NoStop}%
\bibitem [{\citenamefont {Ali-Haïmoud}\ \emph {et~al.}(2017)\citenamefont
  {Ali-Haïmoud}, \citenamefont {Kovetz},\ and\ \citenamefont
  {Kamionkowski}}]{AKK}%
  \BibitemOpen
  \bibfield  {author} {\bibinfo {author} {\bibfnamefont {Y.}~\bibnamefont
  {Ali-Haïmoud}}, \bibinfo {author} {\bibfnamefont {E.~D.}\ \bibnamefont
  {Kovetz}},\ and\ \bibinfo {author} {\bibfnamefont {M.}~\bibnamefont
  {Kamionkowski}},\ }\bibfield  {title} {\bibinfo {title} {Merger rate of
  primordial black-hole binaries},\ }\bibfield  {journal} {\bibinfo  {journal}
  {Physical Review D}\ }\textbf {\bibinfo {volume} {96}},\ \href
  {https://doi.org/10.1103/physrevd.96.123523} {10.1103/physrevd.96.123523}
  (\bibinfo {year} {2017})\BibitemShut {NoStop}%
\bibitem [{Note3()}]{Note3}%
  \BibitemOpen
  \bibinfo {note} {Among the alternative scenarios let us mention the scenario
  of the formation of the bound pairs in galactic cusps at relatively late
  time, see e.g. \cite {Bird_16, Fakhry_21}. Note, however, that this scenario
  seems to require a large mass fraction of PBHs $f\sim 1$, which may be
  refuted by the constraints on $f$ in this mass range, see e.g \cite
  {Carr_21_constr}.}\BibitemShut {Stop}%
\bibitem [{\citenamefont {Mack}\ \emph {et~al.}(2007)\citenamefont {Mack},
  \citenamefont {Ostriker},\ and\ \citenamefont {Ricotti}}]{MOR_2007}%
  \BibitemOpen
  \bibfield  {author} {\bibinfo {author} {\bibfnamefont {K.~J.}\ \bibnamefont
  {Mack}}, \bibinfo {author} {\bibfnamefont {J.~P.}\ \bibnamefont {Ostriker}},\
  and\ \bibinfo {author} {\bibfnamefont {M.}~\bibnamefont {Ricotti}},\
  }\bibfield  {title} {\bibinfo {title} {Growth of structure seeded by
  primordial black holes},\ }\href {https://doi.org/10.1086/518998} {\bibfield
  {journal} {\bibinfo  {journal} {The Astrophysical Journal}\ }\textbf
  {\bibinfo {volume} {665}},\ \bibinfo {pages} {1277–1287} (\bibinfo {year}
  {2007})}\BibitemShut {NoStop}%
\bibitem [{\citenamefont {{Eroshenko}}(2016)}]{Eroshenko_16}%
  \BibitemOpen
  \bibfield  {author} {\bibinfo {author} {\bibfnamefont {Y.~N.}\ \bibnamefont
  {{Eroshenko}}},\ }\bibfield  {title} {\bibinfo {title} {{Dark matter density
  spikes around primordial black holes}},\ }\href
  {https://doi.org/10.1134/S1063773716060013} {\bibfield  {journal} {\bibinfo
  {journal} {Astronomy Letters}\ }\textbf {\bibinfo {volume} {42}},\ \bibinfo
  {pages} {347} (\bibinfo {year} {2016})},\ \Eprint
  {https://arxiv.org/abs/1607.00612} {arXiv:1607.00612 [astro-ph.HE]}
  \BibitemShut {NoStop}%
\bibitem [{\citenamefont {Kavanagh}\ \emph {et~al.}(2018)\citenamefont
  {Kavanagh}, \citenamefont {Gaggero},\ and\ \citenamefont
  {Bertone}}]{kavanagh}%
  \BibitemOpen
  \bibfield  {author} {\bibinfo {author} {\bibfnamefont {B.~J.}\ \bibnamefont
  {Kavanagh}}, \bibinfo {author} {\bibfnamefont {D.}~\bibnamefont {Gaggero}},\
  and\ \bibinfo {author} {\bibfnamefont {G.}~\bibnamefont {Bertone}},\
  }\bibfield  {title} {\bibinfo {title} {Merger rate of a subdominant
  population of primordial black holes},\ }\bibfield  {journal} {\bibinfo
  {journal} {Physical Review D}\ }\textbf {\bibinfo {volume} {98}},\ \href
  {https://doi.org/10.1103/physrevd.98.023536} {10.1103/physrevd.98.023536}
  (\bibinfo {year} {2018})\BibitemShut {NoStop}%
\bibitem [{Note4()}]{Note4}%
  \BibitemOpen
  \bibinfo {note} {By a binary PBH we understand later on a gravitationally
  bound pair of PBHs. We use both these terms interchangeably
  below.}\BibitemShut {Stop}%
\bibitem [{Note5()}]{Note5}%
  \BibitemOpen
  \bibinfo {note} {Note that the energy is not conserved initially, when tidal
  forces determined by expansion of the Universe dominate over gravity of the
  binary. Nonetheless, we use the usual Keplerian definitions of binding energy
  and semimajor axis at all times.}\BibitemShut {Stop}%
\bibitem [{\citenamefont {{Heggie}}(1975)}]{1975MNRAS.173..729H}%
  \BibitemOpen
  \bibfield  {author} {\bibinfo {author} {\bibfnamefont {D.~C.}\ \bibnamefont
  {{Heggie}}},\ }\bibfield  {title} {\bibinfo {title} {{Binary evolution in
  stellar dynamics.}},\ }\href {https://doi.org/10.1093/mnras/173.3.729}
  {\bibfield  {journal} {\bibinfo  {journal} {Monthly Notices of the Royal
  Astronomical Society}\ }\textbf {\bibinfo {volume} {173}},\ \bibinfo {pages}
  {729} (\bibinfo {year} {1975})}\BibitemShut {NoStop}%
\bibitem [{Note6()}]{Note6}%
  \BibitemOpen
  \bibinfo {note} {Note that one can also, in principle, include the dynamic
  friction term in this equation. However, it can be shown to be small during
  the radiation dominated stage due to the effect discussed in \cite
  {Shukhman_1982}.}\BibitemShut {Stop}%
\bibitem [{\citenamefont {{Portegies Zwart}}\ \emph {et~al.}(2009)\citenamefont
  {{Portegies Zwart}}, \citenamefont {{McMillan}}, \citenamefont {{Harfst}},
  \citenamefont {{Groen}}, \citenamefont {{Fujii}}, \citenamefont
  {{Nuall{\'a}in}}, \citenamefont {{Glebbeek}}, \citenamefont {{Heggie}},
  \citenamefont {{Lombardi}}, \citenamefont {{Hut}}, \citenamefont {{Angelou}},
  \citenamefont {{Banerjee}}, \citenamefont {{Belkus}}, \citenamefont
  {{Fragos}}, \citenamefont {{Fregeau}}, \citenamefont {{Gaburov}},
  \citenamefont {{Izzard}}, \citenamefont {{Juri{\'c}}}, \citenamefont
  {{Justham}}, \citenamefont {{Sottoriva}}, \citenamefont {{Teuben}},
  \citenamefont {{van Bever}}, \citenamefont {{Yaron}},\ and\ \citenamefont
  {{Zemp}}}]{amuse1}%
  \BibitemOpen
  \bibfield  {author} {\bibinfo {author} {\bibfnamefont {S.}~\bibnamefont
  {{Portegies Zwart}}}, \bibinfo {author} {\bibfnamefont {S.}~\bibnamefont
  {{McMillan}}}, \bibinfo {author} {\bibfnamefont {S.}~\bibnamefont
  {{Harfst}}}, \bibinfo {author} {\bibfnamefont {D.}~\bibnamefont {{Groen}}},
  \bibinfo {author} {\bibfnamefont {M.}~\bibnamefont {{Fujii}}}, \bibinfo
  {author} {\bibfnamefont {B.~{\'O}.}\ \bibnamefont {{Nuall{\'a}in}}}, \bibinfo
  {author} {\bibfnamefont {E.}~\bibnamefont {{Glebbeek}}}, \bibinfo {author}
  {\bibfnamefont {D.}~\bibnamefont {{Heggie}}}, \bibinfo {author}
  {\bibfnamefont {J.}~\bibnamefont {{Lombardi}}}, \bibinfo {author}
  {\bibfnamefont {P.}~\bibnamefont {{Hut}}}, \bibinfo {author} {\bibfnamefont
  {V.}~\bibnamefont {{Angelou}}}, \bibinfo {author} {\bibfnamefont
  {S.}~\bibnamefont {{Banerjee}}}, \bibinfo {author} {\bibfnamefont
  {H.}~\bibnamefont {{Belkus}}}, \bibinfo {author} {\bibfnamefont
  {T.}~\bibnamefont {{Fragos}}}, \bibinfo {author} {\bibfnamefont
  {J.}~\bibnamefont {{Fregeau}}}, \bibinfo {author} {\bibfnamefont
  {E.}~\bibnamefont {{Gaburov}}}, \bibinfo {author} {\bibfnamefont
  {R.}~\bibnamefont {{Izzard}}}, \bibinfo {author} {\bibfnamefont
  {M.}~\bibnamefont {{Juri{\'c}}}}, \bibinfo {author} {\bibfnamefont
  {S.}~\bibnamefont {{Justham}}}, \bibinfo {author} {\bibfnamefont
  {A.}~\bibnamefont {{Sottoriva}}}, \bibinfo {author} {\bibfnamefont
  {P.}~\bibnamefont {{Teuben}}}, \bibinfo {author} {\bibfnamefont
  {J.}~\bibnamefont {{van Bever}}}, \bibinfo {author} {\bibfnamefont
  {O.}~\bibnamefont {{Yaron}}},\ and\ \bibinfo {author} {\bibfnamefont
  {M.}~\bibnamefont {{Zemp}}},\ }\bibfield  {title} {\bibinfo {title} {{A
  multiphysics and multiscale software environment for modeling astrophysical
  systems}},\ }\href {https://doi.org/10.1016/j.newast.2008.10.006} {\bibfield
  {journal} {\bibinfo  {journal} {New Astronomy}\ }\textbf {\bibinfo {volume}
  {14}},\ \bibinfo {pages} {369} (\bibinfo {year} {2009})},\ \Eprint
  {https://arxiv.org/abs/0807.1996} {arXiv:0807.1996 [astro-ph]} \BibitemShut
  {NoStop}%
\bibitem [{\citenamefont {van Elteren}\ \emph {et~al.}(2014)\citenamefont {van
  Elteren}, \citenamefont {Pelupessy},\ and\ \citenamefont {Zwart}}]{amuse2}%
  \BibitemOpen
  \bibfield  {author} {\bibinfo {author} {\bibfnamefont {A.}~\bibnamefont {van
  Elteren}}, \bibinfo {author} {\bibfnamefont {I.}~\bibnamefont {Pelupessy}},\
  and\ \bibinfo {author} {\bibfnamefont {S.~P.}\ \bibnamefont {Zwart}},\
  }\bibfield  {title} {\bibinfo {title} {Multi-scale and multi-domain
  computational astrophysics},\ }\href {https://doi.org/10.1098/rsta.2013.0385}
  {\bibfield  {journal} {\bibinfo  {journal} {Philosophical Transactions of the
  Royal Society A: Mathematical, Physical and Engineering Sciences}\ }\textbf
  {\bibinfo {volume} {372}},\ \bibinfo {pages} {20130385} (\bibinfo {year}
  {2014})}\BibitemShut {NoStop}%
\bibitem [{\citenamefont {{Gott}}(1975)}]{1975ApJ...201..296G}%
  \BibitemOpen
  \bibfield  {author} {\bibinfo {author} {\bibfnamefont {I.}~\bibnamefont
  {{Gott}}, \bibfnamefont {J.~Richard}},\ }\bibfield  {title} {\bibinfo {title}
  {{On the Formation of Elliptical Galaxies}},\ }\href
  {https://doi.org/10.1086/153887} {\bibfield  {journal} {\bibinfo  {journal}
  {\apj}\ }\textbf {\bibinfo {volume} {201}},\ \bibinfo {pages} {296} (\bibinfo
  {year} {1975})}\BibitemShut {NoStop}%
\bibitem [{\citenamefont {{Gunn}}(1977)}]{1977ApJ...218..592G}%
  \BibitemOpen
  \bibfield  {author} {\bibinfo {author} {\bibfnamefont {J.~E.}\ \bibnamefont
  {{Gunn}}},\ }\bibfield  {title} {\bibinfo {title} {{Massive galactic halos.
  I. Formation and evolution.}},\ }\href {https://doi.org/10.1086/155715}
  {\bibfield  {journal} {\bibinfo  {journal} {The Astrophysical Journal}\
  }\textbf {\bibinfo {volume} {218}},\ \bibinfo {pages} {592} (\bibinfo {year}
  {1977})}\BibitemShut {NoStop}%
\bibitem [{\citenamefont {{Bertschinger}}(1985)}]{1985ApJS...58...39B}%
  \BibitemOpen
  \bibfield  {author} {\bibinfo {author} {\bibfnamefont {E.}~\bibnamefont
  {{Bertschinger}}},\ }\bibfield  {title} {\bibinfo {title} {{Self-similar
  secondary infall and accretion in an Einstein-de Sitter universe}},\ }\href
  {https://doi.org/10.1086/191028} {\bibfield  {journal} {\bibinfo  {journal}
  {The Astrophysical Journal Supplement Series}\ }\textbf {\bibinfo {volume}
  {58}},\ \bibinfo {pages} {39} (\bibinfo {year} {1985})}\BibitemShut {NoStop}%
\bibitem [{Note7()}]{Note7}%
  \BibitemOpen
  \bibinfo {note} {Note that this result is different from that obtained in
  \cite {MOR_2007}, who claimed that $\rho \propto r^{-3}$}\BibitemShut
  {NoStop}%
\bibitem [{\citenamefont {Sesana}\ \emph {et~al.}(2006)\citenamefont {Sesana},
  \citenamefont {Haardt},\ and\ \citenamefont {Madau}}]{sesana}%
  \BibitemOpen
  \bibfield  {author} {\bibinfo {author} {\bibfnamefont {A.}~\bibnamefont
  {Sesana}}, \bibinfo {author} {\bibfnamefont {F.}~\bibnamefont {Haardt}},\
  and\ \bibinfo {author} {\bibfnamefont {P.}~\bibnamefont {Madau}},\ }\bibfield
   {title} {\bibinfo {title} {Interaction of massive black hole binaries with
  their stellar environment. i. ejection of hypervelocity stars},\ }\href
  {https://doi.org/10.1086/507596} {\bibfield  {journal} {\bibinfo  {journal}
  {The Astrophysical Journal}\ }\textbf {\bibinfo {volume} {651}},\ \bibinfo
  {pages} {392–400} (\bibinfo {year} {2006})}\BibitemShut {NoStop}%
\bibitem [{\citenamefont {{Polyachenko}}\ and\ \citenamefont
  {{Shukhman}}(2017)}]{2017MNRAS.470.2190P}%
  \BibitemOpen
  \bibfield  {author} {\bibinfo {author} {\bibfnamefont {E.~V.}\ \bibnamefont
  {{Polyachenko}}}\ and\ \bibinfo {author} {\bibfnamefont {I.~G.}\ \bibnamefont
  {{Shukhman}}},\ }\bibfield  {title} {\bibinfo {title} {{Radial orbit
  instability in systems of highly eccentric orbits: Antonov problem
  reviewed}},\ }\href {https://doi.org/10.1093/mnras/stx1317} {\bibfield
  {journal} {\bibinfo  {journal} {Monthly Notices of the Royal Astronomical
  Society}\ }\textbf {\bibinfo {volume} {470}},\ \bibinfo {pages} {2190}
  (\bibinfo {year} {2017})},\ \Eprint {https://arxiv.org/abs/1705.09150}
  {arXiv:1705.09150 [astro-ph.GA]} \BibitemShut {NoStop}%
\bibitem [{Note8()}]{Note8}%
  \BibitemOpen
  \bibinfo {note} {Note that the value of $M_*$ in their Fig. 5, which is the
  closest to our value $M_*=1$, is $M_*=3/5$. This shouldn't, however,
  influence the agreement in any significant way}\BibitemShut {NoStop}%
\bibitem [{\citenamefont {{De Luca}}\ \emph
  {et~al.}(2020{\natexlab{a}})\citenamefont {{De Luca}}, \citenamefont
  {{Franciolini}}, \citenamefont {{Pani}},\ and\ \citenamefont
  {{Riotto}}}]{DeLuca_20}%
  \BibitemOpen
  \bibfield  {author} {\bibinfo {author} {\bibfnamefont {V.}~\bibnamefont {{De
  Luca}}}, \bibinfo {author} {\bibfnamefont {G.}~\bibnamefont {{Franciolini}}},
  \bibinfo {author} {\bibfnamefont {P.}~\bibnamefont {{Pani}}},\ and\ \bibinfo
  {author} {\bibfnamefont {A.}~\bibnamefont {{Riotto}}},\ }\bibfield  {title}
  {\bibinfo {title} {{Constraints on primordial black holes: The importance of
  accretion}},\ }\href {https://doi.org/10.1103/PhysRevD.102.043505} {\bibfield
   {journal} {\bibinfo  {journal} {Physical Review D}\ }\textbf {\bibinfo
  {volume} {102}},\ \bibinfo {eid} {043505} (\bibinfo {year}
  {2020}{\natexlab{a}})},\ \Eprint {https://arxiv.org/abs/2003.12589}
  {arXiv:2003.12589 [astro-ph.CO]} \BibitemShut {NoStop}%
\bibitem [{\citenamefont {{Hasinger}}(2020)}]{Hasinger_20}%
  \BibitemOpen
  \bibfield  {author} {\bibinfo {author} {\bibfnamefont {G.}~\bibnamefont
  {{Hasinger}}},\ }\bibfield  {title} {\bibinfo {title} {{Illuminating the dark
  ages: cosmic backgrounds from accretion onto primordial black hole dark
  matter}},\ }\href {https://doi.org/10.1088/1475-7516/2020/07/022} {\bibfield
  {journal} {\bibinfo  {journal} {Journal of Cosmology and Astroparticle
  Physics}\ }\textbf {\bibinfo {volume} {2020}}\bibfield  {number} {\bibinfo
  {number} { (7)},\ \bibinfo {eid} {022}},\ }\Eprint
  {https://arxiv.org/abs/2003.05150} {arXiv:2003.05150 [astro-ph.CO]}
  \BibitemShut {NoStop}%
\bibitem [{\citenamefont {{De Luca}}\ \emph
  {et~al.}(2020{\natexlab{b}})\citenamefont {{De Luca}}, \citenamefont
  {{Franciolini}}, \citenamefont {{Pani}},\ and\ \citenamefont
  {{Riotto}}}]{deluca20}%
  \BibitemOpen
  \bibfield  {author} {\bibinfo {author} {\bibfnamefont {V.}~\bibnamefont {{De
  Luca}}}, \bibinfo {author} {\bibfnamefont {G.}~\bibnamefont {{Franciolini}}},
  \bibinfo {author} {\bibfnamefont {P.}~\bibnamefont {{Pani}}},\ and\ \bibinfo
  {author} {\bibfnamefont {A.}~\bibnamefont {{Riotto}}},\ }\bibfield  {title}
  {\bibinfo {title} {{Primordial black holes confront LIGO/Virgo data: current
  situation}},\ }\href {https://doi.org/10.1088/1475-7516/2020/06/044}
  {\bibfield  {journal} {\bibinfo  {journal} {\jcap}\ }\textbf {\bibinfo
  {volume} {2020}},\ \bibinfo {eid} {044} (\bibinfo {year}
  {2020}{\natexlab{b}})},\ \Eprint {https://arxiv.org/abs/2005.05641}
  {arXiv:2005.05641 [astro-ph.CO]} \BibitemShut {NoStop}%
\bibitem [{\citenamefont {Nitz}\ \emph {et~al.}(2019)\citenamefont {Nitz},
  \citenamefont {Capano}, \citenamefont {Nielsen}, \citenamefont {Reyes},
  \citenamefont {White}, \citenamefont {Brown},\ and\ \citenamefont
  {Krishnan}}]{astro_bh}%
  \BibitemOpen
  \bibfield  {author} {\bibinfo {author} {\bibfnamefont {A.~H.}\ \bibnamefont
  {Nitz}}, \bibinfo {author} {\bibfnamefont {C.}~\bibnamefont {Capano}},
  \bibinfo {author} {\bibfnamefont {A.~B.}\ \bibnamefont {Nielsen}}, \bibinfo
  {author} {\bibfnamefont {S.}~\bibnamefont {Reyes}}, \bibinfo {author}
  {\bibfnamefont {R.}~\bibnamefont {White}}, \bibinfo {author} {\bibfnamefont
  {D.~A.}\ \bibnamefont {Brown}},\ and\ \bibinfo {author} {\bibfnamefont
  {B.}~\bibnamefont {Krishnan}},\ }\bibfield  {title} {\bibinfo {title} {1-ogc:
  The first open gravitational-wave catalog of binary mergers from analysis of
  public advanced ligo data},\ }\href
  {https://doi.org/10.3847/1538-4357/ab0108} {\bibfield  {journal} {\bibinfo
  {journal} {The Astrophysical Journal}\ }\textbf {\bibinfo {volume} {872}},\
  \bibinfo {pages} {195} (\bibinfo {year} {2019})}\BibitemShut {NoStop}%
\bibitem [{\citenamefont {Zumalac\'arregui}\ and\ \citenamefont
  {Seljak}(2018)}]{constr_war1}%
  \BibitemOpen
  \bibfield  {author} {\bibinfo {author} {\bibfnamefont {M.}~\bibnamefont
  {Zumalac\'arregui}}\ and\ \bibinfo {author} {\bibfnamefont {U.~c.~v.}\
  \bibnamefont {Seljak}},\ }\bibfield  {title} {\bibinfo {title} {Limits on
  stellar-mass compact objects as dark matter from gravitational lensing of
  type ia supernovae},\ }\href {https://doi.org/10.1103/PhysRevLett.121.141101}
  {\bibfield  {journal} {\bibinfo  {journal} {Phys. Rev. Lett.}\ }\textbf
  {\bibinfo {volume} {121}},\ \bibinfo {pages} {141101} (\bibinfo {year}
  {2018})}\BibitemShut {NoStop}%
\bibitem [{\citenamefont {{Garcia-Bellido}}\ \emph {et~al.}(2017)\citenamefont
  {{Garcia-Bellido}}, \citenamefont {{Clesse}},\ and\ \citenamefont
  {{Fleury}}}]{constr_war2}%
  \BibitemOpen
  \bibfield  {author} {\bibinfo {author} {\bibfnamefont {J.}~\bibnamefont
  {{Garcia-Bellido}}}, \bibinfo {author} {\bibfnamefont {S.}~\bibnamefont
  {{Clesse}}},\ and\ \bibinfo {author} {\bibfnamefont {P.}~\bibnamefont
  {{Fleury}}},\ }\bibfield  {title} {\bibinfo {title} {{LIGO Lo(g)Normal MACHO:
  Primordial Black Holes survive SN lensing constraints}},\ }\href@noop {}
  {\bibfield  {journal} {\bibinfo  {journal} {arXiv e-prints}\ ,\ \bibinfo
  {eid} {arXiv:1712.06574}} (\bibinfo {year} {2017})},\ \Eprint
  {https://arxiv.org/abs/1712.06574} {arXiv:1712.06574 [astro-ph.CO]}
  \BibitemShut {NoStop}%
\bibitem [{\citenamefont {{Bird}}\ \emph {et~al.}(2016)\citenamefont {{Bird}},
  \citenamefont {{Cholis}}, \citenamefont {{Mu{\~n}oz}}, \citenamefont
  {{Ali-Ha{\"\i}moud}}, \citenamefont {{Kamionkowski}}, \citenamefont
  {{Kovetz}}, \citenamefont {{Raccanelli}},\ and\ \citenamefont
  {{Riess}}}]{Bird_16}%
  \BibitemOpen
  \bibfield  {author} {\bibinfo {author} {\bibfnamefont {S.}~\bibnamefont
  {{Bird}}}, \bibinfo {author} {\bibfnamefont {I.}~\bibnamefont {{Cholis}}},
  \bibinfo {author} {\bibfnamefont {J.~B.}\ \bibnamefont {{Mu{\~n}oz}}},
  \bibinfo {author} {\bibfnamefont {Y.}~\bibnamefont {{Ali-Ha{\"\i}moud}}},
  \bibinfo {author} {\bibfnamefont {M.}~\bibnamefont {{Kamionkowski}}},
  \bibinfo {author} {\bibfnamefont {E.~D.}\ \bibnamefont {{Kovetz}}}, \bibinfo
  {author} {\bibfnamefont {A.}~\bibnamefont {{Raccanelli}}},\ and\ \bibinfo
  {author} {\bibfnamefont {A.~G.}\ \bibnamefont {{Riess}}},\ }\bibfield
  {title} {\bibinfo {title} {{Did LIGO Detect Dark Matter?}},\ }\href
  {https://doi.org/10.1103/PhysRevLett.116.201301} {\bibfield  {journal}
  {\bibinfo  {journal} {Physical Review Letters}\ }\textbf {\bibinfo {volume}
  {116}},\ \bibinfo {eid} {201301} (\bibinfo {year} {2016})},\ \Eprint
  {https://arxiv.org/abs/1603.00464} {arXiv:1603.00464 [astro-ph.CO]}
  \BibitemShut {NoStop}%
\bibitem [{\citenamefont {{Fakhry}}\ \emph {et~al.}(2021)\citenamefont
  {{Fakhry}}, \citenamefont {{Firouzjaee}},\ and\ \citenamefont
  {{Farhoudi}}}]{Fakhry_21}%
  \BibitemOpen
  \bibfield  {author} {\bibinfo {author} {\bibfnamefont {S.}~\bibnamefont
  {{Fakhry}}}, \bibinfo {author} {\bibfnamefont {J.~T.}\ \bibnamefont
  {{Firouzjaee}}},\ and\ \bibinfo {author} {\bibfnamefont {M.}~\bibnamefont
  {{Farhoudi}}},\ }\bibfield  {title} {\bibinfo {title} {{Primordial black hole
  merger rate in ellipsoidal-collapse dark matter halo models}},\ }\href
  {https://doi.org/10.1103/PhysRevD.103.123014} {\bibfield  {journal} {\bibinfo
   {journal} {Physical Review D}\ }\textbf {\bibinfo {volume} {103}},\ \bibinfo
  {eid} {123014} (\bibinfo {year} {2021})},\ \Eprint
  {https://arxiv.org/abs/2012.03211} {arXiv:2012.03211 [astro-ph.CO]}
  \BibitemShut {NoStop}%
\bibitem [{\citenamefont {{Carr}}\ \emph {et~al.}(2021)\citenamefont {{Carr}},
  \citenamefont {{Kohri}}, \citenamefont {{Sendouda}},\ and\ \citenamefont
  {{Yokoyama}}}]{Carr_21_constr}%
  \BibitemOpen
  \bibfield  {author} {\bibinfo {author} {\bibfnamefont {B.}~\bibnamefont
  {{Carr}}}, \bibinfo {author} {\bibfnamefont {K.}~\bibnamefont {{Kohri}}},
  \bibinfo {author} {\bibfnamefont {Y.}~\bibnamefont {{Sendouda}}},\ and\
  \bibinfo {author} {\bibfnamefont {J.}~\bibnamefont {{Yokoyama}}},\ }\bibfield
   {title} {\bibinfo {title} {{Constraints on primordial black holes}},\ }\href
  {https://doi.org/10.1088/1361-6633/ac1e31} {\bibfield  {journal} {\bibinfo
  {journal} {Reports on Progress in Physics}\ }\textbf {\bibinfo {volume}
  {84}},\ \bibinfo {eid} {116902} (\bibinfo {year} {2021})},\ \Eprint
  {https://arxiv.org/abs/2002.12778} {arXiv:2002.12778 [astro-ph.CO]}
  \BibitemShut {NoStop}%
\bibitem [{\citenamefont {{Bisnovatyi-Kogan}}\ and\ \citenamefont
  {{Shukhman}}(1982)}]{Shukhman_1982}%
  \BibitemOpen
  \bibfield  {author} {\bibinfo {author} {\bibfnamefont {G.~S.}\ \bibnamefont
  {{Bisnovatyi-Kogan}}}\ and\ \bibinfo {author} {\bibfnamefont {I.~G.}\
  \bibnamefont {{Shukhman}}},\ }\bibfield  {title} {\bibinfo {title} {{Particle
  collisions in an expanding universe.}},\ }\href@noop {} {\bibfield  {journal}
  {\bibinfo  {journal} {Zhurnal Eksperimentalnoi i Teoreticheskoi Fiziki}\
  }\textbf {\bibinfo {volume} {82}},\ \bibinfo {pages} {3} (\bibinfo {year}
  {1982})}\BibitemShut {NoStop}%
\end{thebibliography}%

\end{document}